\documentclass[aps,prd,twocolumn,nofootinbib,superscriptaddress,floatfix]{revtex4-2}

\usepackage[T1]{fontenc}
\usepackage[utf8]{inputenc}
\usepackage{amsmath,amssymb,amsfonts,bm}
\usepackage{graphicx}
\usepackage{xcolor}
\usepackage{booktabs}
\usepackage{array}
\usepackage{multirow}
\usepackage{hyperref}
\usepackage{physics}
\usepackage{placeins} 
\usepackage{fnpct} 
\usepackage{tikz} \usetikzlibrary{calc}
\usepackage{comment}
\usepackage{ifpdf}
\usepackage{slashed}
\usepackage[normalem]{ulem}
\usepackage{placeins}
\usepackage{cancel}

\hypersetup{
  colorlinks=true,
  linkcolor=blue,
  citecolor=blue,
  urlcolor=blue
}

\newcommand{\rh}{r_{\rm h}}
\newcommand{\lamG}{\lambda_G}

\newcommand{\maybeinclude}[2][]{%
  \IfFileExists{#2}{\includegraphics[#1]{#2}}{%
  \fbox{\parbox{0.92\linewidth}{\centering Figure placeholder: \texttt{#2}}}}}

\definecolor{lime}{HTML}{A6CE39}
\newcommand{\orcidicon}{%
	\begin{tikzpicture}
	\draw[lime, fill=lime] (0,0)
	circle [radius=0.16]
	node[white] {{\fontfamily{qag}\selectfont \tiny ID}};
	\draw[white, fill=white] (-0.0625,0.095)
	circle [radius=0.007];
	\end{tikzpicture}   \hspace{-2mm}
}

\newcommand\orcidFaical{{\href{https://orcid.org/0000-0002-2977-0821}{\orcidicon}}}
\newcommand\orcidWijdane{{\href{https://orcid.org/0000-0001-6103-1079}{\orcidicon}}}
\newcommand\orcidHasan{{\href{https://orcid.org/0000-0001-7408-0910}{\orcidicon}}}
\newcommand\orcidKarima{{\href{https://orcid.org/0009-0009-4463-2353}{\orcidicon}}}

\begin{document}

\title{\textcolor{brown}{ Accretion and Neutrino-Motivated Thermal Diagnostics of
Thermodynamically Reconstructed Black Holes from
Three-Parameter Generalized Entropy}
}

\author{F. Barzi\orcidFaical\!\!}
\email{faical.barzi@edu.uiz.ac.ma}
\affiliation{\small LPTHE, Physics Department, Faculty of Sciences,  Ibnou Zohr University, Agadir, Morocco.}
\affiliation{\small CRMEF, Regional Center for Education and Training Professions, Marrakesh, Morocco. }

\author{W.  El Hadri \orcidWijdane}
\email{w.elhadri@uiz.ac.ma}
\affiliation{\small LPTHE, Physics Department, Faculty of Sciences,  Ibnou Zohr University, Agadir, Morocco.}
\affiliation{\small ENSIASD, Taroudant, Ibnou Zohr University, Agadir, Morocco.}

\author{H. El Moumni \orcidHasan}	
\email{h.elmoumni@uiz.ac.ma (Corresponding author)}

\author{K.  Masmar \orcidKarima}
\email{k.masmar@uiz.ac.ma}
\affiliation{\small LPTHE, Physics Department, Faculty of Sciences,  Ibnou Zohr University, Agadir, Morocco.}

\author{S. Mazzou}
\email{safaamazzou23@gmail.com}
\affiliation{\small LPTHE, Physics Department, Faculty of Sciences, Ibnou Zohr University, Agadir, Morocco.}

\date{\today}

\begin{abstract}

We study accretion and thermal diagnostics motivated by neutrino
processes for a static black hole reconstructed from a
three-parameter generalized entropy.
The construction is thermodynamic by design: the entropy deformation is
not mapped to a radial-coordinate redefinition or to a prescribed
Reissner--Nordstr\"om-like correction. Instead, the generalized entropy
fixes the horizon response factor $\Xi_h=dS_G/dS|_{S_h}$. The effective
exterior geometry is then reconstructed by requiring its
surface-gravity temperature to reproduce the generalized
thermodynamic temperature. As a result, the metric preserves the
horizon area and the Schwarzschild asymptotics, and reduces smoothly to
Schwarzschild when $\Xi_h\to1$.

The entropy response is encoded in
$\lambda_G=1/\Xi_{\rm h}-1$, whereas the exterior radial dependence
requires an additional localization prescription.  The detailed
accretion analysis is performed for the power-law family
$h(x)=x^{-p}$, with $p\geq2$.  Within this reconstruction family,
positive $\lambda_G$ moves the photon sphere and ISCO inward,
decreases the critical shadow scale, raises the radiative efficiency,
and concentrates the energy release toward the inner disk, while
negative $\lambda_G$ produces the opposite trend.  A complementary
comparison with exponential and rational localization functions shows
that this qualitative ordering is preserved for the profiles tested,
although the magnitude of the strong-field shifts remains
profile dependent.

Finally, the neutrino-motivated thermal sector is modeled
conservatively using dimensionless temperature moments instead of a
full neutrino-dominated accretion flow or annihilation-deposition
calculation.
The hierarchy among the $ n=4$, $n=6$, and $n=9$ moments reveals that entropy-induced disk deformation becomes increasingly apparent with higher temperature exponents. 
Thus, within the adopted reconstruction framework, generalized entropy
fixes the horizon response, while the exterior localization prescription
determines how this response is propagated into compact-orbit and
thin-disk observables.

\end{abstract}

\maketitle

\tableofcontents

\section{Introduction}

Black-hole thermodynamics is one of the few places where gravity,
quantum field theory and information theory meet in a quantitative way. More broadly, the relation between horizon thermodynamics and
gravitational dynamics has led to the interpretation of the Einstein
equations themselves as an equation of state
\cite{Jacobson:1995ab}.
The mechanical laws of stationary black holes already suggest a
thermodynamic reading of surface gravity and horizon area
\cite{BardeenCarterHawking1973}. Subsequently, Bekenstein's area law
and Hawking's radiation calculation fixed the entropy and temperature
of the Schwarzschild black hole in the form now known as the
Bekenstein--Hawking prescription
\cite{Bekenstein1973,Hawking1975}. This prescription is remarkably
robust. Nevertheless, it is not immune to conceptual pressure. Quantum
corrections, non-additive information measures, spacetime
microstructure and phenomenological entropy deformations have all
motivated alternatives to the simple area law.

Accordingly, several entropy functionals have appeared in gravitational
and cosmological settings. Rényi entropy \cite{Renyi:1961}, Tsallis
entropy \cite{Tsallis:1987eu}, Sharma--Mittal entropy
\cite{SharmaMittal1975}, Kaniadakis entropy \cite{Kaniadakis2001}, and
Barrow's fractal-area proposal \cite{Barrow:2020tzx} are among the most
frequently used examples. In the black-hole context, Rényi statistics
has proved especially useful because it can generate nontrivial
thermodynamic behavior even in asymptotically flat spacetimes. It has
been used, in particular, to analyze critical phenomena,
Hawking--Page-type universality, phase-equilibrium structure,
thermodynamic topology and chaotic response in charged-flat and
dilatonic black-hole systems
\cite{Promsiri:2020jga,ElMoumniMasmarMazzou2022,BarziElMoumni2022,
Promsiri:2021hhv,BarziElMoumniMasmar2023,
BarziElMoumniMasmar2024JHEAp,Nakarachinda:2021jxd,
BarziElMoumniMasmar2024NPB}.

More recently, Nojiri, Odintsov and Faraoni proposed a generalized
entropy family that contains several of these possibilities as limiting
or special cases. More importantly, they emphasized a point that is
crucial for black holes: once the entropy is modified, temperature and
thermodynamic energy cannot be left unchanged without checking the
first law
\cite{Nojiri:2022aof,NojiriOdintsovFaraoniAstrophysics2022}. This observation provides the starting point of the present work.

Recent analyses have stressed that, if the Schwarzschild geometry and
the ADM mass are both kept fixed, the Hawking temperature and the
Bekenstein--Hawking entropy remain uniquely selected
\cite{Elizalde:2025iku}. An alternative route is to derive
the modified field equations implied by the generalized entropy, which
can lead to an area-dependent effective gravitational coupling
\cite{Lu:2024ppa}. These viewpoints make it essential to
separate the thermodynamic input from the assumptions used to
reconstruct the exterior geometry.

However, the issue is not merely semantic. In a number of recent
applications, an entropy deformation has been translated directly into
a deformation of the radial coordinate or of the exterior metric. Such
a step is not justified without additional geometric assumptions.
Indeed, the entropy is a horizon quantity; by itself it fixes the
thermodynamic response at the horizon, not the full radial dependence
of the geometry. Therefore, if a metric is introduced, one must clearly
separate the part of the geometry fixed by thermodynamics from the part
that remains an effective modeling choice. Otherwise, apparent changes
in the ISCO or in disk observables may reflect a coordinate convention
rather than a physical modification of the spacetime.

In this work, we propose a minimal thermodynamic reconstruction. The
metric is written in terms of a radial coordinate $r$ for which the
two-spheres have area $A=4\pi r^2$. The generalized entropy fixes the
ratio between the generalized temperature and the Hawking temperature.
We then impose this temperature on the surface gravity of an effective
static metric. As a result, the entropy response defines the
one-parameter horizon deformation
$
\lamG=\frac{1}{\Xi_{\rm h}}-1,
$
where $\Xi_{\rm h}=dS_G/dS|_{S_{\rm h}}$. At the same time, the full
radial profile is not claimed to be unique. We parametrize it by an
integer $p\ge2$, which controls how rapidly the correction decays away
from the horizon. 
The representative values $p=2,3,4$ are used to quantify the residual
dependence within the power-law localization family.  Since the
thermodynamic matching does not uniquely determine the exterior radial
profile, we also test distinct exponential and rational localization
functions below.

The astrophysical motivation comes from accretion. Thin disks around
compact objects remain among the most direct probes of strong-field
gravity. The classical Shakura--Sunyaev disk
\cite{Shakura:1973} and the relativistic
Novikov--Thorne/Page--Thorne construction
\cite{Novikov:1973,Page:1974,Thorne:1974} provide a controlled
setting in which the ISCO, the orbital energy and the radiative
efficiency enter directly. In addition, the photon sphere fixes the
leading shadow scale, while redshift and Doppler effects shape the
observed spectrum \cite{Cunningham:1975,Luminet:1979,Belhaj:2022vte,Chakhchi:2024tzo}. Consequently, the
goal is to isolate and track the leading changes induced by the
entropy-driven reconstruction through a minimal set of compact and
disk-based observables. This is not intended to replace full ray
tracing or GRMHD modeling, but rather to identify the clean
strong-field imprint of the thermodynamic deformation.

A closely related study considered Novikov--Thorne accretion and
neutrino-pair annihilation for a Barrow-modified black hole constructed
through the direct replacement $r\rightarrow r^{1+\Delta/2}$
\cite{Shi:2025msw}. The present construction differs at this point:
the generalized entropy fixes only the horizon response, while the
exterior radial profile is introduced separately as an effective
localization choice.

A second motivation comes from hyperaccretion and neutrino physics.
Neutrino-cooled disks and neutrino--antineutrino annihilation have long
been studied as possible ingredients of gamma-ray-burst engines
\cite{PophamWoosleyFryer1999,DiMatteoPernaNarayan2002,
ChenBeloborodov2007,ZalameaBeloborodov2011,Liu2017}. A complete
treatment would require vertical disk structure, composition, neutrino
opacity, electron degeneracy and relativistic transport, none of which
are resolved here. Instead, we introduce dimensionless temperature
moments designed to measure how the entropy-induced modification of the
disk temperature propagates into quantities with increasingly strong
thermal sensitivity. The exponents $n=4,6,9$ are used as benchmarks for
thermal, charged-current/Urca-like, and pair-process or
annihilation-like temperature dependences. Thus, the neutrino sector is
treated as a conservative thermal probe of the inner disk, not as a
self-consistent neutrino-dominated accretion-flow model or a physical
annihilation-deposition calculation.

The paper is organized as follows. Section~\ref{sec:entropy}
introduces the three-parameter entropy and the horizon thermodynamics.
Section~\ref{sec:geometry} presents the thermodynamic reconstruction of
the effective metric and its curvature diagnostics. Section~\ref{sec:geodesics}
derives the circular-orbit structure, photon sphere, ISCO and compact
observables. Section~\ref{sec:disk} develops the disk flux,
temperature, energy-at-infinity luminosity and redshifted spectral
proxy. Section~\ref{sec:neutrinos} introduces the neutrino-sensitive
temperature moments. Section~\ref{sec:numerical_results} then discusses
the representative strong-field signatures of the thermodynamic
reconstruction. Finally, Sec.~\ref{sec:conclusion} is devoted to the conclusion and future perspectives.

\section{Three-parameter entropy and horizon thermodynamics}
\label{sec:entropy}

Black-hole thermodynamics provides one of the clearest indications that gravity, quantum theory and statistical mechanics are not independent frameworks. For a stationary black hole, the Bekenstein--Hawking entropy is proportional to the horizon area,
\begin{equation}
S=\frac{A}{4G}.
\end{equation}
For a spherical horizon of radius $\rh$, one has
\begin{equation}
S_{\rm h}
=
\frac{\pi \rh^2}{G}.
\label{eq:Sh}
\end{equation}
For the Schwarzschild geometry, the mass-radius relation and the Hawking temperature are
\begin{equation}
M=
\frac{\rh}{2G},
\qquad
T_H=
\frac{1}{4\pi \rh}.
\label{eq:SchwMassTemp}
\end{equation}
In the standard Bekenstein--Hawking description these quantities satisfy the first law,
\begin{equation}
T_H\,dS=dM.
\end{equation}

The area law is expected to be modified once quantum, statistical or microscopic gravitational effects are taken into account. Black holes are also systems governed by long-range gravitational interactions, and this makes them natural candidates for generalized, non-additive or non-extensive entropic descriptions. Several deformations have been studied in this context, including Tsallis entropy, R\'enyi entropy, Sharma--Mittal entropy, Barrow entropy, Kaniadakis entropy and loop-quantum-gravity inspired non-extensive entropies; see, for example, Refs.~\cite{Tsallis:1987eu,Renyi:1961,Czinner:2015eyk,SayahianJahromi:2018irq,Barrow:2020tzx,Kaniadakis:2005zk,Tsallis:2012js,Majhi:2017zao,Nojiri:2022aof,Nojiri:2022sfd}. These proposals have different motivations, but they share a common goal: to replace the strict area law by a controlled entropy deformation that remains positive, increases with the Bekenstein--Hawking entropy, vanishes when the latter vanishes, and admits the standard area law as a limiting case.

In this work we use the three-parameter entropy introduced in Ref.~\cite{Nojiri:2022aof},
\begin{equation}
S_G(\alpha,\beta,\gamma)
=
\frac{1}{\gamma}
\left[
\left(
1+\frac{\alpha}{\beta}S
\right)^\beta
-1
\right],
\label{eq:SG}
\end{equation}
where $S$ denotes the standard Bekenstein--Hawking entropy. In the numerical and analytical discussion below we work on the branch
\begin{equation}
\alpha>0,
\qquad
\beta>0,
\qquad
\gamma>0,
\end{equation}
for which $S_G>0$ and $dS_G/dS>0$ whenever $S>0$. The deformation is therefore introduced at the level of the entropy functional, while the geometric horizon area remains
\begin{equation}
A_{\rm h}=4\pi \rh^2.
\end{equation}
We do not replace the horizon radius by a deformed one. This point is essential for the construction developed in the next section. 
The entropy \eqref{eq:SG} contains several known entropic forms as limiting cases. The Bekenstein--Hawking entropy is recovered, for example, for
\begin{equation}\label{eqq8}
\beta=1,
\qquad
\gamma=\alpha,
\end{equation}
for which
\begin{equation}\label{eqq9}
S_G=S,
\qquad
\frac{dS_G}{dS}=1.
\end{equation}
When $\gamma=\alpha$, Eq.\eqref{eq:SG} reduces to a Sharma--Mittal-type expression with the undeformed Bekenstein--Hawking entropy as its base entropy. For a suitable scaling of $\gamma$ and large $\alpha$, one obtains power-law entropies of the form $S^\beta$, including Tsallis-type and Barrow-type deformations for appropriate choices of $\beta$. The R\'enyi limit is slightly more delicate because it requires a correlated scaling. Setting
\begin{equation}
q=\frac{\alpha}{\beta},
\end{equation}
then taking $\beta\to0$ at fixed $q$ with $\gamma=\alpha=q\beta$, one obtains
\begin{equation}
S_G
\longrightarrow
\frac{1}{q}
\ln(1+qS),
\end{equation}
which is the usual R\'enyi form. 
 This limit is not only formal. It connects the present three-parameter entropy to earlier applications of Rényi black-hole thermodynamics, where the nonextensive parameter modifies the thermal response and phase structure while the Bekenstein--Hawking area remains the reference entropy variable~\cite{Hirunsirisawat:2022fsb,BarziElMoumniMasmar2023}.

Thus the relevant limits are not obtained by sending the three parameters independently to special values; they arise from correlated scalings that keep the entropy normalization finite.

The quantity that enters the first law is not only $S_G$, but rather its response to a variation of the Bekenstein--Hawking entropy. We therefore define
\begin{equation}
\Xi(S)
\equiv
\frac{dS_G}{dS}.
\label{eq:XiDef}
\end{equation}
For Eq.\eqref{eq:SG}, this gives
\begin{equation}
\Xi(S)
=
\frac{\alpha}{\gamma}
\left(
1+\frac{\alpha}{\beta}S
\right)^{\beta-1}.
\label{eq:Xi}
\end{equation}
At the black-hole horizon,
\begin{equation}
\Xi_{\rm h}
=
\left.
\frac{dS_G}{dS}
\right|_{S=S_{\rm h}}
=
\frac{\alpha}{\gamma}
\left(
1+\frac{\alpha\pi \rh^2}{\beta G}
\right)^{\beta-1}.
\label{eq:Xih}
\end{equation}

We keep $M$ as the thermodynamic energy. The generalized temperature is therefore not a free choice. Since
\begin{equation}
dS_G=\Xi_{\rm h}\,dS,
\end{equation}
the first law requires
\begin{equation}
T_G\,dS_G=dM.
\end{equation}
Using the standard relation $T_H dS=dM$, one obtains
\begin{equation}
T_G
=
\frac{T_H}{\Xi_{\rm h}}.
\label{eq:TG}
\end{equation}
This expression is the thermodynamic origin of the deformation used in the rest of the paper. Indeed,
\begin{align}
T_G\frac{dS_G}{d\rh}
&=
\frac{T_H}{\Xi_{\rm h}}\,
\Xi_{\rm h}\frac{dS}{d\rh}
\nonumber\\
&=
T_H\frac{dS}{d\rh}
=
\frac{dM}{d\rh}
=
\frac{1}{2G}.
\label{eq:first_law_check}
\end{align}
Thus the first law is recovered identically. In the present construction, the entropy deformation is not interpreted as a direct deformation of the radial coordinate. It is interpreted as a deformation of the thermodynamic response of the horizon. This is why $\Xi_{\rm h}$, rather than $S_G$ alone, is the quantity that drives the effective reconstruction.

It is useful to introduce the dimensionless parameter
\begin{equation}
\lambda_G
=
\frac{1}{\Xi_{\rm h}}-1.
\label{eq:lambda}
\end{equation}
In terms of $\lambda_G$, the generalized temperature reads
\begin{equation}
T_G
=
(1+\lambda_G)T_H.
\label{eq:TGlambda}
\end{equation}
Hence $\lambda_G$ measures the fractional change of the horizon temperature induced by the generalized entropy. The condition
\begin{equation}
\lambda_G>-1
\end{equation}
ensures a positive horizon temperature.  The local Schwarzschild-response condition at a given horizon is
\begin{equation}
\Xi_h = 1, \qquad \lambda_G = 0.
\end{equation}
At that horizon, this condition implies
$T_G=T_H$ and therefore removes the thermodynamic deformation
entering the reconstructed geometry. It should not, however, be
identified with the global Bekenstein--Hawking entropy limit:
$\Xi_h=1$ fixes only the local derivative
$dS_G/dS|_{S_h}$ and does not in general imply the functional
identity $S_G(S)=S$. The genuine Bekenstein--Hawking entropy
limit is obtained when $S_G=S$ for all $S$, which for the
three-parameter entropy considered here is realized by
$\beta=1$ and $\gamma=\alpha$, as shown in Eqs.~\eqref{eqq8} and \eqref{eqq9}.

For numerical work, $\lambda_G$ is the most transparent parameter because it directly controls the surface gravity of the reconstructed geometry. This does not mean that the three-parameter entropy has been abandoned. For fixed $\alpha$, $\beta$ and $\rh$, one may prescribe a value of $\lambda_G$ and reconstruct the corresponding $\gamma$. From Eq.\eqref{eq:Xih}, one obtains
\begin{equation}
\gamma
=
\alpha(1+\lambda_G)
\left(
1+\frac{\alpha S_{\rm h}}{\beta}
\right)^{\beta-1}.
\label{eq:gamma_mapping}
\end{equation}
 Eq.\eqref{eq:gamma_mapping} should be understood as a calibration at a fixed
horizon entropy.  Conversely, if the entropy parameters
$(\alpha,\beta,\gamma)$ are regarded as fixed parameters of a given
entropy theory, the horizon-response parameter is generally
mass dependent:
\begin{equation}
1+\lambda_G(M)
=
\frac{\gamma}{\alpha}
\left[
1+\frac{\alpha S_h(M)}{\beta}
\right]^{1-\beta}.
\label{eq:lambda_mass}
\end{equation}
For a Schwarzschild black hole,
\begin{equation}
\frac{S_h}{k_B}
=
\frac{4\pi G M^2}{\hbar c}
\simeq
1.049\times10^{77}
\left(\frac{M}{M_\odot}\right)^2 .
\label{eq:BH_entropy_physical}
\end{equation}
Thus a fixed entropy theory does not, in general, assign the same
value of $\lambda_G$ to black holes of different masses.  In the
large-entropy regime, $\alpha S_h/\beta\gg1$,
\begin{equation}
1+\lambda_G\propto S_h^{\,1-\beta}
\propto M^{2(1-\beta)} .
\end{equation}
Consequently, $1+\lambda_G$ grows with mass for $0<\beta<1$,
is mass independent for $\beta=1$, and decreases with mass for
$\beta>1$.

As an explicit illustrative mapping at an astrophysically
relevant reference mass, consider a Schwarzschild black hole of mass
$M_0=10\,M_\odot$, for which
$S_0/k_B\simeq1.049\times10^{79}$.  Taking
$\beta=1/2$ and choosing $\alpha S_0/\beta=1$ gives
$\alpha\simeq4.77\times10^{-80}$ in the present entropy
normalization.  Equation~(23) then yields
\begin{equation}
\gamma\simeq
\begin{cases}
4.04\times10^{-80}, & \lambda_G=+0.2,\\
2.70\times10^{-80}, & \lambda_G=-0.2.
\end{cases}
\footnote{ Equivalently, the relevant dimensionless combinations are
$\alpha S_0=0.5$ and $\gamma S_0\simeq0.424$ or $0.283$ for
$\lambda_G=+0.2$ and $-0.2$, respectively.}
\end{equation}
This reference choice is not inferred from observations; it is used
only to demonstrate explicitly that the benchmark values
$\lambda_G=\pm0.2$ can be realized with positive entropy parameters
at a realistic black-hole entropy scale. 
Thus the representative values used in the strong-field analysis
can be realized by explicit positive values of
$(\alpha,\beta,\gamma)$ at an astrophysically relevant horizon
entropy.

Once $(\alpha,\beta,\gamma)$ are fixed by such a calibration,
$\lambda_G$ should not be readjusted when the black-hole mass is
changed.  For the choice $\alpha S_0/\beta=1$, its subsequent
mass dependence is
\begin{equation}
1+\lambda_G(M)
=
(1+\lambda_{G,0})
\left[
\frac{1+(M/M_0)^2}{2}
\right]^{1-\beta}.
\label{eq:lambda_mass_calibrated}
\end{equation}
Accordingly, the values $\lambda_G=\pm0.2$ used below should be
viewed as representative horizon-response benchmarks at a fixed
reference entropy, rather than as mass-independent predictions of
a single fixed entropy theory.

\begin{figure}[!ht]
\centering
\includegraphics[width=\columnwidth]{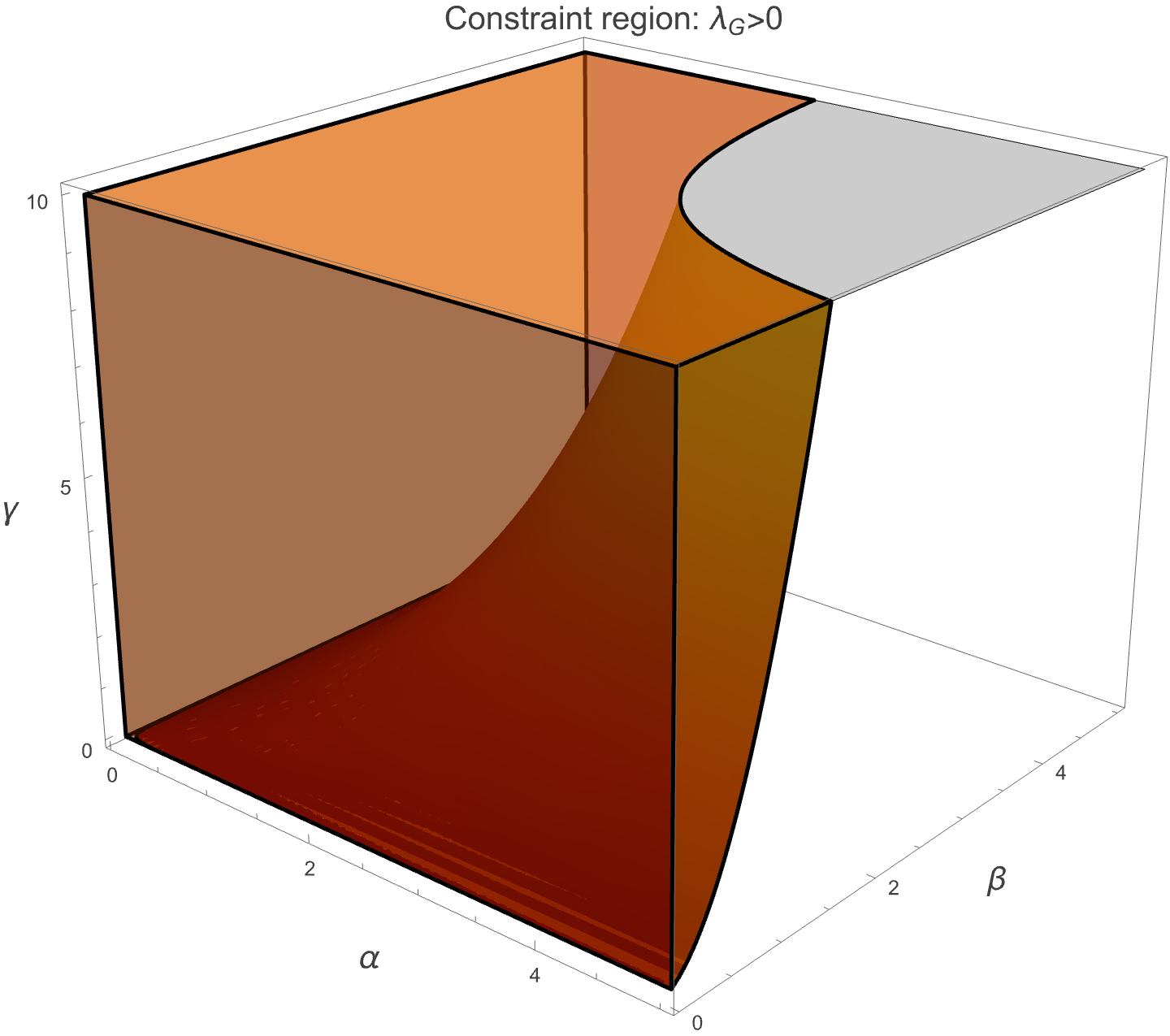}
\caption{\footnotesize{\it 
Illustrative normalized constraint region in the three-parameter
entropy space $(\alpha,\beta,\gamma)$ corresponding to
$\lambda_G>0$ at $S_h=1$.  This normalized plot is intended only
to display the parameter-space sectors; an explicit astrophysical
calibration at $M_0=10\,M_\odot$ is given in the text. The orange volume represents the sector in which the generalized temperature satisfies $T_G>T_H$. The boundary surface is defined by $\lambda_G=0$, or equivalently $\gamma=\gamma_{\rm crit}(\alpha,\beta)$, with $\gamma_{\rm crit}=\alpha(1+\alpha S_h/\beta)^{\beta-1}$. The region $\lambda_G<0$ is not excluded; it corresponds to $T_G<T_H$ as long as $\lambda_G>-1$.
}}
\label{fig:lambdaG_positive_constraint}
\end{figure}

Therefore, scanning $\lambda_G$ amounts to scanning the particular
combination of entropy parameters that fixes the horizon temperature.
For fixed $S_h$, the sign of $\lambda_G$ defines two sectors in the
entropy-parameter space. The critical surface
\begin{equation}
\gamma_{\rm crit}(\alpha,\beta)
=
\alpha
\left(
1+\frac{\alpha S_h}{\beta}
\right)^{\beta-1}
\label{eq:gamma_crit}
\end{equation}
corresponds to $\lambda_G=0$. Values $\gamma>\gamma_{\rm crit}$ give
$\lambda_G>0$ and therefore $T_G>T_H$, whereas
$0<\gamma<\gamma_{\rm crit}$ gives $-1<\lambda_G<0$ and therefore
$T_G<T_H$. Figure~\ref{fig:lambdaG_positive_constraint} illustrates
the sector $\lambda_G>0$ in the three-parameter entropy space.

The sign of $\lambda_G$ has a direct physical meaning. If $\lambda_G>0$, then
\begin{equation}
T_G>T_H,
\end{equation}
and the generalized entropy enhances the horizon temperature. If $\lambda_G<0$, then
\begin{equation}
T_G<T_H,
\end{equation}
and the generalized entropy lowers it. 
 Within the exterior reconstruction adopted below, $\lambda_G$
parametrizes the thermodynamic response at the horizon. The general
analysis in Sec.~IV will show, however, that both the direction and
magnitude of the photon-sphere and ISCO shifts also depend on the
radial localization profile. 
 The purpose of the next section is to translate the thermodynamic input \eqref{eq:TG} into a minimal effective geometry whose surface gravity reproduces $T_G$, while preserving the physical horizon radius and the ADM mass.

\section{Thermodynamic reconstruction of the effective metric}
\label{sec:geometry}

\subsection{Metric ansatz and thermodynamic matching}

We now translate the thermodynamic input of Sec.~\ref{sec:entropy} into a minimal effective geometry. We consider a static and spherically symmetric line element
\begin{equation}
ds^2 = -f_G(r)dt^2 + \frac{dr^2}{f_G(r)} + r^2 d\Omega^2,
\label{eq:metric}
\end{equation}
where the two-spheres have area $A(r)=4\pi r^2$. The horizon is kept at $r=r_h$, and we adopt the following blackening  function profile
\begin{equation}
f_G(r) =\left( 1-\frac{r_h}{r} \right)\left[1+\lambda_G\left(\frac{r_h}{r}
\right)^p \right],\qquad p\geq2.
\label{eq:fG}
\end{equation}
This form is chosen to satisfy four minimal requirements. First, $f_G(r_h)=0$, so the horizon position is not shifted. Second, the correction does not modify the asymptotic Schwarzschild mass term. Indeed,
\begin{equation}
f_G(r) = 1-\frac{r_h}{r} + \lambda_G\frac{r_h^p}{r^p}- \lambda_G\frac{r_h^{p+1}}{r^{p+1}},
\label{eq:asymptotic_fG}
\end{equation}
and for $p\geq2$ there is no new contribution at order $1/r$. The ADM mass is therefore still given by
\begin{equation}
M_{\rm ADM}=\frac{r_h}{2G}.
\label{eq:ADM_mass}
\end{equation}
Third, the Schwarzschild geometry is recovered smoothly when $\lambda_G\to0$. Fourth, the surface-gravity temperature must reproduce the generalized temperature obtained from the entropy response.

The parameter $p$ is not an entropy parameter.  More generally, the
same thermodynamic matching at the horizon can be realized by
\begin{equation}
f_h(x)=\left(1-\frac{1}{x}\right)
      \left[1+\lambda_G h(x)\right],
\qquad h(1)=1 ,
\label{eq:generic_profile}
\end{equation}
where $x=r/r_h$.  Indeed, the condition $h(1)=1$ implies
$f_h'(1)=1+\lambda_G$ and therefore reproduces the same
surface-gravity temperature fixed by the entropy response.  To
preserve the Schwarzschild ADM mass, we additionally require
$h(x)=o(x^{-1})$ as $x\rightarrow\infty$.\\ 
Equation~\eqref{eq:generic_profile} makes explicit that $\lambda_G$
and $h(x)$ represent two distinct ingredients of the effective
construction.  The former is fixed by the generalized entropy through
the horizon response, whereas the latter specifies an additional
phenomenological completion of the exterior geometry.  The power-law
choice used for the detailed accretion analysis,
\begin{equation}
h_{\rm P}(x)=x^{-p},\qquad p\geq2 ,
\label{eq:power_profile}
\end{equation}
is therefore one admissible localization family rather than a unique
consequence of the generalized entropy.  The choices $p=2,3,4$
quantify the residual dependence within this family.  In Sec.~IV we
also compare the reference $p=2$ case with exponential and rational
profiles having the same value and first derivative at the horizon.

We emphasize that this construction is phenomenological and effective.
The thermodynamic matching fixes the horizon surface gravity, but no
underlying covariant action or modified field equation is assumed from
which both the generalized entropy and the exterior metric follow
simultaneously. Accordingly, the reconstructed metric should not be
interpreted as a unique geometry derived from $S_G$, but as an effective
exterior completion constrained by the horizon thermodynamics.

For the power-law family and $\lambda_G>-1$, the second factor in
Eq.\eqref{eq:fG}  remains positive for $r\geq r_h$, so no additional outer
horizon is introduced.  The negative-$\lambda_G$ branch nevertheless has a distinct interior
structure. For $-1<\lambda_G<0$, the second factor of the power-law
metric vanishes at the specific location
\begin{equation}
x_- = (-\lambda_G)^{1/p}<1,
\end{equation}
which defines an additional inner horizon. As
$\lambda_G\rightarrow-1^{+}$, this root approaches the outer horizon
$x=1$ and the surface gravity tends to zero. The boundary value
$\lambda_G=-1$, excluded from the finite-temperature parameter range
used here, therefore corresponds to a degenerate-horizon limit.  
The surface-gravity temperature is
\begin{equation}
T_{\rm geo}
=
\frac{f_G'(r_h)}{4\pi}
=
\frac{1+\lambda_G}{4\pi r_h}.
\label{eq:Tgeo}
\end{equation}
Using the definition $\lambda_G=1/\Xi_h-1$, this becomes
\begin{equation}
T_{\rm geo}
=
\frac{1}{4\pi r_h\Xi_h}
=
T_G.
\label{eq:Tgeo_TG}
\end{equation}
Thus the metric reconstruction is thermodynamically tied to the generalized entropy through the horizon response factor $\Xi_h$. The local Schwarzschild-response limit of the reconstructed geometry is
\begin{equation}
\Xi_h\rightarrow1,
\qquad
\lambda_G\rightarrow0,
\qquad
f_G(r)\rightarrow1-\frac{r_h}{r}.
\label{eq:classical_limit_metric}
\end{equation}

\subsection{Curvature and effective source}

The reconstructed metric is not a vacuum Einstein solution when $\lambda_G\neq0$. It is therefore useful to characterize the geometry through an effective stress tensor,
\begin{equation}
G^\mu{}_\nu
=
8\pi G\,T^\mu{}_{\nu,\rm eff},
\label{eq:Einstein_eff}
\end{equation}
with
\begin{equation}
T^\mu{}_{\nu,\rm eff}
=
{\rm diag}
[-\rho_{\rm eff},p_{r,\rm eff},p_{t,\rm eff},p_{t,\rm eff}].
\label{eq:Teff_def}
\end{equation}
For compactness, the following expressions are displayed in geometrized units $G=1$:
\begin{equation}
\rho_{\rm eff}
=
\frac{
\lambda_G(r_h/r)^p
\left[(p-1)r-pr_h\right]
}
{8\pi r^3},
\label{eq:rho_eff}
\end{equation}
\begin{equation}
p_{r,\rm eff}
=
-\rho_{\rm eff},
\label{eq:pr_eff}
\end{equation}
and
\begin{equation}
p_{t,\rm eff}
=
\frac{
\lambda_G p(r_h/r)^p
\left[(p-1)r-(p+1)r_h\right]
}
{16\pi r^3}.
\label{eq:pt_eff}
\end{equation}
All these quantities vanish in the Schwarzschild limit $\lambda_G=0$ and decay at large radius. The radial null energy condition is saturated,
\begin{equation}
\rho_{\rm eff}+p_{r,\rm eff}=0,
\label{eq:radial_NEC}
\end{equation}
whereas the tangential null energy condition and the strong energy condition depend on $\lambda_G$, $p$ and $r$. We do not interpret $T^\mu{}_{\nu,\rm eff}$ as ordinary matter. It is an effective source that parametrizes the non-vacuum geometry needed to encode the entropy-induced modification of the surface gravity. It may therefore violate the usual pointwise energy conditions in some regions, as expected for an effective thermodynamic correction.  At the event horizon these statements become particularly transparent:
\begin{equation}
\rho_{\rm eff}(r_h)
=-\frac{\lambda_G}{8\pi r_h^2},
\qquad
p_{r,\rm eff}(r_h)
=\frac{\lambda_G}{8\pi r_h^2},
\qquad
p_{t,\rm eff}(r_h)
=-\frac{p\lambda_G}{8\pi r_h^2}.
\end{equation}
Hence the radial null energy condition remains saturated, whereas for
$\lambda_G>0$ the effective energy density is negative and the
tangential null energy condition is violated at the horizon. The
strong energy condition is likewise violated there for
$\lambda_G>0$. These violations do not represent exotic material
matter in the present framework; they characterize the effective
stress tensor associated with the phenomenologically reconstructed
non-vacuum geometry.

One curvature scalar that makes the departure from Schwarzschild explicit is
\begin{equation}
R
=
\lambda_G(p-1)
\left(
\frac{r_h}{r}
\right)^p
\frac{
p r_h-(p-2)r
}
{r^3}.
\label{eq:Ricci_scalar}
\end{equation}
It vanishes for $\lambda_G=0$ and remains finite at $r=r_h$ for finite $\lambda_G$ and $p\geq2$. The quadratic invariants show the same behavior. In particular, in the Schwarzschild limit one recovers
\begin{eqnarray}
R&=&0,
\qquad
R_{\mu\nu}R^{\mu\nu}=0,\\
\nonumber
K
&=&
R_{\mu\nu\rho\sigma}R^{\mu\nu\rho\sigma}
=
\frac{12r_h^2}{r^6}
=
\frac{48M^2}{r^6}.
\label{eq:Schw_invariants}
\end{eqnarray}

\section{Circular geodesics, photon sphere and ISCO}
\label{sec:geodesics}

We now study the circular-orbit structure of the thermodynamically reconstructed geometry. It is convenient to introduce the dimensionless radial variable
\begin{equation}
x=\frac{r}{r_h},
\label{eq:x_def}
\end{equation}
so that the metric function becomes
\begin{equation}
f_G(x)
=
\left(1-\frac{1}{x}\right)
\left(1+\frac{\lambda_G}{x^p}\right).
\label{eq:fG_x}
\end{equation}
Throughout this section, primes denote derivatives with respect to $x$. 
Since $x$ is dimensionless, we write the orbital frequency and angular momentum in dimensionless form,
\begin{equation}
\widetilde{\Omega}_\phi
=
r_h\Omega_\phi,
\qquad
\widetilde L
=
\frac{L}{r_h}.
\label{eq:dimensionless_orbital_quantities}
\end{equation}
The physical orbital frequency is recovered from $\Omega_\phi=\widetilde{\Omega}_\phi/r_h$. In the disk formulas below, the same dimensionless convention is used unless dimensions are explicitly restored.

For equatorial timelike geodesics, $\theta=\pi/2$, the conserved energy and angular momentum per unit rest mass are
\begin{equation}
E
=
f_G(r)\dot t,
\qquad
L
=
r^2\dot\phi,
\label{eq:EL_def}
\end{equation}
where the dot denotes differentiation with respect to proper time. The normalization condition $u^\mu u_\mu=-1$ gives
\begin{equation}
\dot r^2
+
V_{\rm eff}(r)
=
E^2,
\label{eq:radial_equation}
\end{equation}
with the effective potential
\begin{equation}
V_{\rm eff}(r)
=
f_G(r)
\left(
1+\frac{L^2}{r^2}
\right).
\label{eq:radial_effective_potential}
\end{equation}
In terms of $x=r/r_h$ and $\widetilde L=L/r_h$, this potential becomes
\begin{equation}
V_{\rm eff}(x)
=
f_G(x)
\left(
1+\frac{\widetilde L^2}{x^2}
\right).
\label{eq:dimensionless_effective_potential}
\end{equation}
Circular timelike orbits satisfy the following constraint 
\begin{equation}
V_{\rm eff}(x)
=
E^2,
\qquad
\frac{dV_{\rm eff}}{dx}
=
0,
\label{eq:circular_orbit_potential_conditions}
\end{equation}
where $\widetilde L$ is held fixed in the radial derivative.

Solving these two conditions gives the standard circular-orbit quantities for a static spherical metric~\cite{Bardeen:1972fi,Chandrasekhar:1983}. The angular frequency is
\begin{equation}
\widetilde{\Omega}_\phi^2(x)
=
\frac{f_G'(x)}{2x}.
\label{eq:Omega_phi}
\end{equation}
It is useful to introduce the following quantity
\begin{equation}
D(x)
=
f_G(x)
-
\frac{x}{2}f_G'(x).
\label{eq:Dx_def}
\end{equation}
The conserved energy and angular momentum along circular timelike orbits are then
\begin{equation}
E(x)
=
\frac{f_G(x)}{\sqrt{D(x)}},
\qquad
\widetilde L(x)
=
\frac{x^2\widetilde{\Omega}_\phi(x)}{\sqrt{D(x)}}.
\label{eq:E_L_circular}
\end{equation}
Physical circular timelike orbits require
\begin{equation}
f_G(x)>0,
\qquad
\widetilde{\Omega}_\phi^2(x)>0,
\qquad
D(x)>0.
\label{eq:circular_orbit_conditions}
\end{equation}
The last condition ensures that $E$ and $\widetilde L$ are real. Its saturation also marks the limiting null circular orbit.

For null geodesics, the radial potential is proportional to $f_G(x)/x^2$. The photon sphere is therefore determined by
\begin{equation}
\frac{d}{dx}
\left[
\frac{f_G(x)}{x^2}
\right]
=
0,
\label{eq:null_circular_condition}
\end{equation}
or equivalently
\begin{equation}
D(x)
=
0,
\qquad
2f_G(x)-x f_G'(x)
=
0.
\label{eq:photon_sphere_eq}
\end{equation}
This explains why the timelike circular-orbit quantities degenerate at the photon sphere.

In the Schwarzschild limit, $\lambda_G=0$, Eq.~\eqref{eq:photon_sphere_eq} gives
\begin{equation}
x_{\rm ph}^{(0)}
=
\frac{3}{2}.
\label{eq:xph_schw}
\end{equation}
For small $\lambda_G$, the photon-sphere position admits the perturbative expansion
\begin{equation}
x_{\rm ph}
=
\frac{3}{2}
-
\lambda_G
\frac{p\,2^{p-2}}{3^p}
+
\mathcal O(\lambda_G^2).
\label{eq:xph_pert}
\end{equation}
Thus, for $p>0$, a positive $\lambda_G$ shifts the photon sphere inward, while a negative $\lambda_G$ shifts it outward.

The corresponding dimensionless critical impact parameter is
\begin{equation}
\widetilde b_{\rm ph}
=
\frac{x_{\rm ph}}{\sqrt{f_G(x_{\rm ph})}}.
\label{eq:bph_def}
\end{equation}
The use of the critical null orbit to define the leading shadow scale follows the classic escape-cone analysis of Synge and the later black-hole imaging literature~\cite{Synge:1966,Luminet:1979,Perlick:2021aok}. The physical shadow radius, understood as the critical impact parameter measured at infinity, is therefore
\begin{equation}
b_{\rm ph}
=
r_h\,\widetilde b_{\rm ph}
=
r_h
\frac{x_{\rm ph}}{\sqrt{f_G(x_{\rm ph})}}.
\label{eq:physical_shadow_radius}
\end{equation}
Moreover, for a distant observer located at distance $\mathcal{D}_{\rm obs}$, the angular shadow radius is
\begin{equation}
\theta_{\rm sh}
=
\frac{b_{\rm ph}}{\mathcal{D}_{\rm obs}}
=
\frac{r_h}{\mathcal{D}_{\rm obs}}
\frac{x_{\rm ph}}{\sqrt{f_G(x_{\rm ph})}}.
\label{eq:angular_shadow_radius}
\end{equation}
The corresponding angular shadow diameter is
\begin{equation}
\theta_{\rm d}
=
2\theta_{\rm sh}.
\label{eq:angular_shadow_diameter}
\end{equation}
This angular scale is the quantity that can be compared with shadow-size measurements once the mass and distance of the source are specified, as in the EHT analyses of M87* and Sgr A*~\cite{EHT:2019M87I,EHT:2022SgrAI,Jusufi:2021fek}.

In the Schwarzschild limit,
\begin{equation}
\widetilde b_{\rm ph}^{(0)}
=
\frac{3\sqrt{3}}{2},
\qquad
b_{\rm ph}^{(0)}
=
\frac{3\sqrt{3}}{2}r_h.
\label{eq:bph_schw}
\end{equation}
Using $r_h=2GM$ in units $c=1$, this becomes
\begin{equation}
b_{\rm ph}^{(0)}
=
3\sqrt{3}\,GM.
\label{eq:shadow_schw_physical}
\end{equation}
At first order in $\lambda_G$, the dimensionless shadow scale changes as
\begin{equation}
\frac{\widetilde b_{\rm ph}}{\widetilde b_{\rm ph}^{(0)}}
=
1
-
\frac{\lambda_G}{2}
\left(\frac{2}{3}\right)^p
+
\mathcal O(\lambda_G^2).
\label{eq:bph_pert}
\end{equation}
Since $r_h$ and $\mathcal{D}_{\rm obs}$ are kept fixed in the present comparison, the same fractional correction applies to the angular shadow radius,
\begin{equation}
\frac{\theta_{\rm sh}}{\theta_{\rm sh}^{(0)}}
=
1
-
\frac{\lambda_G}{2}
\left(\frac{2}{3}\right)^p
+
\mathcal O(\lambda_G^2).
\label{eq:theta_shadow_pert}
\end{equation}
Thus a positive $\lambda_G$ reduces the leading shadow scale, whereas a negative $\lambda_G$ increases it. This quantity should be understood as the critical-orbit shadow scale. A full image, including the brightness distribution, lensing transfer function and inclination-dependent optical appearance, requires ray tracing and is beyond the scope of the present work.

The innermost stable circular orbit is obtained from the marginal-stability condition
\begin{equation}
\frac{d\widetilde L^2}{dx}
=
0.
\label{eq:isco_dL}
\end{equation}
Within the metric \eqref{eq:fG_x}, this condition can be written in the compact form
\begin{equation}
3f_G f_G'
+
x f_G f_G''
-
2x(f_G')^2
=
0.
\label{eq:isco_eq}
\end{equation}
In the Schwarzschild limit, Eq.~\eqref{eq:isco_eq} gives
\begin{equation}
x_{\rm ISCO}^{(0)}
=
3.
\label{eq:isco_schw}
\end{equation}
This corresponds to
\begin{equation}
r_{\rm ISCO}^{(0)}
=
3r_h
=
6GM.
\label{eq:isco_schw_physical}
\end{equation}
For small $\lambda_G$, the ISCO position is
\begin{equation}
x_{\rm ISCO}
=
3
+
\lambda_G
\frac{4p(1-p)}{3^p}
+
\mathcal O(\lambda_G^2).
\label{eq:isco_pert}
\end{equation}
Since $p>1$ in the present reconstruction, the coefficient of $\lambda_G$ is negative. Therefore,
\begin{equation}
\lambda_G>0
\quad\Longrightarrow\quad
x_{\rm ISCO}<3,
\label{eq:isco_positive_lambda}
\end{equation}
whereas
\begin{equation}
-1<\lambda_G<0
\quad\Longrightarrow\quad
x_{\rm ISCO}>3.
\label{eq:isco_negative_lambda}
\end{equation}
This analytic result explains the numerical trend: a positive $\lambda_G$ moves the stable inner edge of the disk closer to the horizon, while a negative $\lambda_G$ pushes it outward.

For an asymptotically flat static spacetime, the conserved energy $E$ is normalized so that a massive particle at rest at infinity has $E=1$. The binding energy per unit rest mass of a circular orbit is therefore
\begin{equation}
E_{\rm bind}(x)
=
1-E(x).
\label{eq:binding_energy}
\end{equation}
At the inner edge of a Novikov--Thorne disk, this binding energy gives the maximum radiative efficiency in the zero-torque approximation. The Novikov--Thorne radiative efficiency is therefore~\cite{Novikov:1973,Page:1974,Thorne:1974}
\begin{equation}
\eta_{\rm NT}
=
E_{\rm bind}(x_{\rm ISCO})
=
1-E(x_{\rm ISCO}).
\label{eq:eta_NT}
\end{equation}
Since $E(x_{\rm ISCO})$ decreases when the ISCO moves inward, a positive $\lambda_G$ increases $\eta_{\rm NT}$, while a negative $\lambda_G$ lowers it. The efficiency therefore measures how the reconstructed geometry changes the energetics of the inner disk.

The radial epicyclic frequency follows from a small radial perturbation of a circular orbit. Epicyclic frequencies are standard probes of orbital stability and are often used in relativistic disk and QPO modeling~\cite{Stella:1998mq,Stella:1999mt,Kato:2008,Stuchlik:2014foa,Ahal:2025zxc, Ahal:2026ksi}. Starting from the effective potential in Eq.~\eqref{eq:radial_effective_potential}, one may write, in coordinate time,
\begin{equation}
\Omega_r^2
=
\frac{f_G(r_0)^2}{2E(r_0)^2}
\left.
\frac{d^2V_{\rm eff}}{dr^2}
\right|_{r=r_0},
\label{eq:Omega_r_from_potential_r}
\end{equation}
where $r_0$ is the radius of the reference circular orbit. After rewriting the result in terms of $x=r/r_h$ and substituting the circular-orbit expressions for $E$ and $\widetilde L$, one obtains
\begin{equation}
\widetilde{\Omega}_r^2
=
\frac{1}{2}
\left[
f_G f_G''
-
2(f_G')^2
+
\frac{3}{x}f_G f_G'
\right].
\label{eq:Omega_r}
\end{equation}
The ISCO is equivalently characterized by
\begin{equation}
\widetilde{\Omega}_r^2(x_{\rm ISCO})
=
0.
\label{eq:Omega_r_ISCO}
\end{equation}
For a static spherical geometry, the vertical epicyclic frequency coincides with the azimuthal frequency,
\begin{equation}
\widetilde{\Omega}_\theta
=
\widetilde{\Omega}_\phi.
\label{eq:Omega_theta}
\end{equation}
In the Schwarzschild limit, Eq.~\eqref{eq:Omega_r} gives
\begin{equation}
\frac{\widetilde{\Omega}_r^2}{\widetilde{\Omega}_\phi^2}
=
1-\frac{3}{x}.
\label{eq:schwarzschild_epicyclic_ratio}
\end{equation}
This ratio vanishes at $x=3$, recovering the Schwarzschild ISCO. This also checks that the epicyclic-frequency criterion and the marginal-stability condition $d\widetilde L^2/dx=0$ identify the same inner stable orbit.

\subsection{Dependence on the exterior localization profile}
\label{subsec:profile_dependence}

The preceding compact-observable results were derived for the
power-law localization $h_{\rm P}(x)=x^{-p}$.  To distinguish the
part fixed by the horizon thermodynamics from that associated with the
exterior completion, consider the more general family in
Eq.~\eqref{eq:generic_profile}.  Writing
$x_{\rm ph}=3/2+\delta x_{\rm ph}$ and expanding the null circular-orbit
condition to first order in $\lambda_G$ gives
\begin{equation}
\delta x_{\rm ph}
=
\frac{3}{8}\lambda_G
h'\!\left(\frac{3}{2}\right).
\label{eq:generic_xph_shift}
\end{equation}
Likewise, expanding the marginal-stability condition around the
Schwarzschild value $x_{\rm ISCO}=3$ yields
\begin{equation}
\delta x_{\rm ISCO}
=
-\lambda_G
\left[
24h'(3)+36h''(3)
\right].
\label{eq:generic_xisco_shift}
\end{equation}
Equations~\eqref{eq:generic_xph_shift} and
\eqref{eq:generic_xisco_shift} show explicitly that the entropy-response
parameter $\lambda_G$ alone does not determine the direction of the
strong-field shifts for an arbitrary exterior completion.  The
photon-sphere shift probes the first derivative of the localization
profile, whereas the ISCO probes both its first and second derivatives.

To test the robustness of the power-law results within physically
comparable localized deformations, we use
\begin{align}
h_{\rm P}(x)&=x^{-2},\\
h_{\rm E}(x)&=e^{-2(x-1)},\\
h_{\rm R}(x)&=
\frac{1}{1+2(x-1)+2(x-1)^2}.
\end{align}
All three profiles satisfy
\begin{equation}
h(1)=1,\qquad h'(1)=-2,\qquad
h(x)=o(x^{-1})\quad (x\rightarrow\infty),
\end{equation}
and are positive and monotonically decreasing for $x\geq1$.  Thus
they reproduce the same horizon temperature, share the same leading
localization slope at the horizon, and preserve the Schwarzschild ADM
mass, while differing in their exterior radial behavior.  Their
first-order coefficients are
\begin{align}
\frac{\delta x_{\rm ph}}{\lambda_G}
&=
-\frac{2}{9},
\quad
-\frac{3}{4e},
\quad
-\frac{6}{25},
\\
\frac{\delta x_{\rm ISCO}}{\lambda_G}
&=
-\frac{8}{9},
\quad
-96e^{-4},
\quad
-\frac{2208}{2197},
\end{align}
for $(h_{\rm P},h_{\rm E},h_{\rm R})$, respectively.

\begin{table*}[!ht]
\centering
\caption{\footnotesize{\it Profile-dependence test for the thermodynamic reconstruction.
The three exterior localization functions satisfy $h(1)=1$ and
$h'(1)=-2$ and preserve the Schwarzschild ADM mass.  The sign pattern
is unchanged for the profiles tested, whereas the magnitude of the
shifts is profile dependent.}}
\label{tab:profile_robustness}
\begin{ruledtabular}
\begin{tabular}{lccccc}
Profile & $\lambda_G$ & $x_{\rm ph}$ & $x_{\rm ISCO}$
& $b_c/b_c^{\rm Schw}$ & $\eta_{\rm NT}$ \\
\hline
$h_{\rm P}(x)=x^{-2}$ & $-0.2$ & 1.54994 & 3.19100 & 1.04602 & 0.05374 \\
$h_{\rm P}(x)=x^{-2}$ & $+0.2$ & 1.46052 & 2.83772 & 0.95725 & 0.06065 \\
$h_{\rm E}(x)=e^{-2(x-1)}$ & $-0.2$ & 1.56100 & 3.31407 & 1.03656 & 0.05322 \\
$h_{\rm E}(x)=e^{-2(x-1)}$ & $+0.2$ & 1.45089 & 2.63718 & 0.96342 & 0.06429 \\
$h_{\rm R}(x)=[1+2(x-1)+2(x-1)^2]^{-1}$ & $-0.2$ & 1.55287 & 3.20391 & 1.04074 & 0.05355 \\
$h_{\rm R}(x)=[1+2(x-1)+2(x-1)^2]^{-1}$ & $+0.2$ & 1.45683 & 2.80503 & 0.96097 & 0.06137 \\
\end{tabular}
\end{ruledtabular}
\end{table*}
The exact numerical comparison in
Table~\ref{tab:profile_robustness} shows that, for the representative
values $\lambda_G=\pm0.2$, the three localization families preserve the
same qualitative ordering found in the original power-law
reconstruction: positive $\lambda_G$ moves the photon sphere and ISCO
inward, reduces the critical shadow scale, and raises the
Novikov--Thorne efficiency, while negative $\lambda_G$ produces the
opposite trend.  The magnitude of these effects, however, is
profile dependent, most noticeably for the ISCO.  We therefore regard
the sign pattern as robust within the localization families tested
here, rather than as a model-independent consequence of generalized
entropy alone.

\section{Thin-disk flux, temperature and redshift proxy}
\label{sec:disk}

\subsection{Novikov--Thorne flux and energy-at-infinity luminosity}
\label{subsec:nt_disk}

The geodesic modifications of Sec.~\ref{sec:geodesics} are now propagated into the thin-disk sector. The accreting gas is regarded as test matter that is minimally coupled to the reconstructed metric; it is assumed that there is no energy or momentum exchange with the effective source that was introduced in Section \ref{sec:geometry}.
We use the Novikov-Thorne description of a stationary, axisymmetric, geometrically thin, and optically thick disk, where the inner edge is at the ISCO~\cite{Novikov:1973,Page:1974,Thorne:1974,Shakura:1973}, and matter moves on nearly circular geodesics. At the inner edge, the typical zero-torque condition is enforced. Thin-disk fluxes and spectra have also been widely used to identify
possible signatures of non-Schwarzschild or modified-gravity
geometries \cite{Pun:2008ae,Ouyang:2025qru,Hu:2025mmp,Cordeiro:2025eox,Ahmed:2025ywb}.

 The Page--Thorne
conservation equations are often quoted in coordinates adapted to the
vertical direction of the disk. When the metric is written in adapted
spherical coordinates, the vertical integration carries the measure
$\sqrt{g_{\theta\theta}}\,d\theta$. The geometric factor entering the
surface flux is consequently $\sqrt{-g/g_{\theta\theta}}$, rather than
the full four-dimensional factor $\sqrt{-g}$. A direct derivation of
this form in spherical coordinates was given in
Ref.~\cite{WuFengChen2024}. In units $c=1$, the physical flux emitted
from one face of the disk is
\begin{equation}
F_{\rm phys}(r)
=
-\frac{\dot M}
{4\pi\sqrt{-g/g_{\theta\theta}}}
\frac{\Omega_{\phi,r}}
{\left(E-\Omega_\phi L\right)^2}
\int_{r_{\rm ISCO}}^{r}
\left(E-\Omega_\phi L\right)
L_{,r'}\,dr' ,
\label{eq:NT_flux_general}
\end{equation}
where $\dot M$ is the rest-mass accretion rate and $E$ and $L$ are the
specific energy and angular momentum of the circular geodesics.

For the considered geometry Eq.\eqref{eq:metric}, the physical flux emitted from one face of the disk becomes
\begin{equation}
F_{\rm phys}(r) = -\frac{\dot M}{4\pi r} \frac{\Omega_{\phi,r}}
{\left(E-\Omega_\phi L\right)^2} \int_{r_{\rm ISCO}}^{r} \left(E-\Omega_\phi L\right) L_{,r'}\,dr' .
\label{eq:NT_flux_spherical}
\end{equation}
Recalling the dimensionless orbital quantities in Eq.~\eqref{eq:dimensionless_orbital_quantities}
and their radial derivatives
\begin{equation}
\Omega_{\phi,r}
= \frac{1}{r_h^2}\widetilde{\Omega}_{\phi,x},
\qquad
L_{,r} =\widetilde L_{,x}.
\label{eq:disk_derivative_scaling}
\end{equation}
The flux may then be written as
\begin{equation}
F_{\rm phys}(r) = \frac{\dot M}{r_h^2} \widetilde F(x;\lambda_G,p),
\label{eq:Fphys_scaling}
\end{equation}
where
\begin{eqnarray}
F(x) &=&
-\frac{1}{4\pi x} \frac{\widetilde{\Omega}_{\phi,x}}{D(x)}
\int_{x_{\rm ISCO}}^{x} \sqrt{D(x')}\,
\widetilde L_{,x'}\,dx'.
\label{eq:flux}
\end{eqnarray}
All orbital quantities in Eq.~\eqref{eq:flux} are evaluated along the
circular geodesics derived in Sec.~\ref{sec:geodesics}. The minus sign
gives a positive outward flux because
$\widetilde{\Omega}_{\phi,x}<0$. The lower integration limit implements
the zero-torque condition at $x_{\rm ISCO}$.


The radial normalization can be tested independently in the weak-field
regime. At large radius,
\begin{equation}
\Omega_\phi
\simeq
\left(\frac{GM}{r^3}\right)^{1/2},
\qquad
L
\simeq
\left(GMr\right)^{1/2},
\qquad
E\simeq1.
\label{eq:newtonian_orbital_limit}
\end{equation}
Substitution into Eq.~\eqref{eq:NT_flux_spherical} yields
\begin{equation}
F_{\rm Newt}(r) =\frac{3GM\dot M}{8\pi r^3} \left[ 1- \left( \frac{r_{\rm in}}{r} \right)^{1/2} \right],
\label{eq:newtonian_flux_limit}
\end{equation}
with $r_{\rm in}=r_{\rm ISCO}$ in the present model. This is the
standard Shakura--Sunyaev thin-disk flux with a zero-torque inner
boundary~\cite{Shakura:1973}. Far from the inner edge,
$r\gg r_{\rm in}$, it reduces to the familiar asymptotic behavior
$F_{\rm Newt}(r)\propto r^{-3}$.  In the dimensionless variables 
the same check reads
\begin{equation}
\widetilde F(x)
\simeq
\frac{3}{16\pi x^3}
\left[
1-
\left(
\frac{x_{\rm ISCO}}{x}
\right)^{1/2}
\right],
\qquad
x\gg x_{\rm ISCO},
\label{eq:newtonian_flux_dimensionless}
\end{equation}
where $GM=r_h/2$ has been used.

The local effective temperature follows from the
Stefan--Boltzmann law,
\begin{equation}
F_{\rm phys}(r)
=
\sigma_{\rm SB}T_{\rm phys}(x)^4.
\label{eq:Stefan_Boltzmann_disk}
\end{equation}
Defining
\begin{equation}
T_0
=
\left(
\frac{\dot M}
{\sigma_{\rm SB}r_h^2}
\right)^{1/4},
\label{eq:T0_disk}
\end{equation}
one obtains
\begin{equation}
T_{\rm phys}(x)
=
T_0\widetilde T(x),
\qquad
\widetilde T(x)
=
\widetilde F(x)^{1/4}.
\label{eq:Tphys_disk}
\end{equation}
The scale $T_0$ contains the dependence on the accretion rate and black-hole size, whereas the dimensionless profile $\widetilde T(x)$ encodes the dependence on $\lambda_G$ and $p$.

$F_{\rm phys}$ is a quantity which is measured locally in the comoving frame of the disk. We use the Page--Thorne energy balance to build up a global quantity associated with the conserved Killing energy. The annulus radiative contribution to the energy carried away to infinity is
\begin{equation}
d\mathcal L_\infty
=
4\pi r\,E(r)\,F_{\rm phys}(r)\,dr,
\label{eq:dLinf_annulus}
\end{equation}
where the factor $4\pi$ accounts for both disk faces. Furthermore, we define the dimensionless energy-at-infinity luminosity by
\begin{equation}
\widetilde{\mathcal L}_\infty
\equiv
\frac{\mathcal L_\infty}{\dot M}
\label{eq:Linf_normalization}
\end{equation}
in units $c=1$. Its radial distribution is
\begin{equation}
\frac{d\widetilde{\mathcal L}_\infty}{d\ln x}
=
4\pi x^2E(x)\widetilde F(x),
\label{eq:dLinfdlnx}
\end{equation}
and the luminosity accumulated up to $x_{\rm out}$ is
\begin{equation}
\widetilde{\mathcal L}_\infty(x_{\rm out})
= \int_{x_{\rm ISCO}}^{x_{\rm out}} 4\pi xE(x)\widetilde F(x)\,dx.
\label{eq:Linf}
\end{equation}
With physical units restored, the corresponding normalization is
$\mathcal L_\infty/(\dot M c^2)$.

For an ideal Novikov--Thorne disk extending to infinity, with zero
torque at the ISCO and neglecting photon capture and returning
radiation, integration of the energy-conservation equation gives
\begin{equation}
\widetilde{\mathcal L}_\infty(\infty)
= 1-E(x_{\rm ISCO})
= \eta_{\rm NT}.
\label{eq:Linf_eta}
\end{equation}
This relation offers a strict global check of the numerical integration and directly connects the flux calculation to the radiative efficiency obtained in Sec.~\ref{sec:geodesics}. When $x_{\rm out}$ is finite, only the radiation produced within that radius is included in Eq.~\eqref{eq:Linf}, and as the outer boundary moves outward, it approaches Eq.~\eqref{eq:Linf_eta}.  
For the numerical evaluation, the Novikov--Thorne flux is tabulated
on a nonuniform radial grid extending to $x_{\max}=200$. The radial
spacing is refined near the ISCO, with $\Delta x=10^{-3}$ over the
first $0.2$ radial units, followed by $\Delta x=10^{-2}$ up to
$x=10$, $\Delta x=0.05$ up to $x=30$, $\Delta x=0.25$ up to
$x=100$, and $\Delta x=1$ at larger radii. Cumulative radial
integrals are evaluated with a composite trapezoidal rule.

Because the convergence of $\widetilde{L}_{\infty}(x_{\rm out})$
toward $\eta_{\rm NT}$ is relatively slow, we evaluate it at
$x_{\rm out}=50,75,100,150,200$ and extrapolate to infinity using
the asymptotic form
$\widetilde{L}_{\infty}(x_{\rm out})
=\widetilde{L}_{\infty}(\infty)+a/x_{\rm out}
+b/x_{\rm out}^{3/2}$.
For all representative cases with $p=2,3,4$ and
$\lambda_G=-0.2,0,0.2$, the extrapolated result satisfies
$\widetilde{L}_{\infty}(\infty)=\eta_{\rm NT}$ with a maximum
fractional residual of $5.9\times10^{-4}$.

An angle-integrated luminosity within the optimal Novikov--Thorne energy budget is represented by the quantity $\widetilde{\mathcal L}_\infty$. It is not yet the luminosity as determined by a specific far-off observer. Eq.~\eqref{eq:Linf} does not include photon capture, gravitational lensing, Doppler beaming, projection effects, or the image-plane transfer function. Below, the directional modulation caused by the leading redshift and Doppler factors is examined independently.

\subsection{Redshift and inclination proxy}
\label{subsec:redshift_proxy}

Along a vacuum null ray, Liouville's theorem implies that the photon
phase-space distribution is conserved. The specific intensity therefore
satisfies
\begin{equation}
\frac{I_\nu}{\nu^3}
=
{\rm invariant},
\label{eq:Liouville_invariant}
\end{equation}
and hence
\begin{equation}
I_{\nu_{\rm obs}}^{\rm obs}
=
g^3
I_{\nu_{\rm obs}/g}^{\rm em},
\qquad
g
=
\frac{\nu_{\rm obs}}{\nu_{\rm em}}.
\label{eq:g_definition}
\end{equation}
A calculation of the observed image or spectrum would require
null-geodesic ray tracing and an image-plane transfer function,
including gravitational lensing and higher-order photon
paths~\cite{Lindquist:1966,Cunningham:1975}. Here our aim is more
limited. We just introduce a no-bending proxy to determine whether the
entropy-induced modifications of the inner disk remain visible after
the leading gravitational and kinematic frequency shifts are included.

For a circular emitter and a static observer at infinity, the redshift
factor is
\begin{equation}
g = \frac{1}
{u^t\left(1-\widetilde{\Omega}_\phi\widetilde b_\phi\right)},
\label{eq:redshift}
\end{equation}
where
\begin{equation}
b_\phi \equiv \frac{k_\phi}{-k_t},
\qquad \widetilde b_\phi = \frac{b_\phi}{r_h}, \text{ and }
u^t =\frac{1}{\sqrt{D(x)}}.
\label{eq:photon_axial_impact_parameter}
\end{equation}
Herein, $b_\phi$ is the axial photon impact parameter.
In the no-bending approximation, the photon trajectory is replaced by
a straight line connecting the emitting point to the distant observer.
The projected axial impact parameter is then modeled as
\begin{equation}
\widetilde b_\phi
\simeq
x\sin i\sin\phi,
\label{eq:b_proxy}
\end{equation}
where $i$ is the observer inclination and $\phi$ is the disk azimuth.
The overall sign depends on the orientation adopted for $\phi$; with
the convention in Eq.~\eqref{eq:b_proxy}, the approaching and receding
sides are distinguished by the sign of $\sin\phi$.

Specific intensities acquire a factor $g^3$. After integration over
frequency, the transformation of the frequency measure introduces one
additional factor, so that bolometric intensities scale as $g^4$.
Assuming locally isotropic emission from the disk surface,
\begin{equation}
I_{\rm em}^{\rm bol}
=
\frac{F_{\rm phys}}{\pi}.
\label{eq:isotropic_bolometric_intensity}
\end{equation}
Replacing the exact image-plane element by the flat projected area
element gives the schematic bolometric flux proxy
\begin{equation}
\widetilde{\mathcal F}_{\rm obs}^{\rm proxy}(i)
\propto
\cos i
\int_{x_{\rm ISCO}}^{x_{\rm out}}dx
\int_0^{2\pi}d\phi\,
x\,g^4(x,\phi,i)\,
\widetilde F(x).
\label{eq:bolometric_proxy}
\end{equation}
The omitted proportionality factor contains the overall geometric
normalization, including the source distance. The factor $\cos i$ is
the Euclidean projection factor in the no-bending approximation and
may be omitted when flux-normalized results are compared at a fixed
inclination. The quantity in Eq.~\eqref{eq:bolometric_proxy} is
directional and should not be identified with the angle-integrated
energy-at-infinity luminosity $\widetilde{\mathcal L}_\infty$ defined
in Eq.~\eqref{eq:Linf}.

A redshifted multicolor-blackbody proxy may similarly be written as
\begin{equation}
\mathcal F_{\nu_{\rm obs}}^{\rm proxy}(i)
\propto
\cos i
\int_{x_{\rm ISCO}}^{x_{\rm out}}dx
\int_0^{2\pi}d\phi\,
x\,g^3(x,\phi,i)\,
B_{\nu_{\rm obs}/g}
\left[
T_{\rm phys}(x)
\right].
\label{eq:mbb_proxy}
\end{equation}
The Planck function is evaluated at the physical disk temperature with the frequency argument being the emitted frequency $\nu_{em} = \frac{\nu_{obs}}{g}$, to be treated as a redshifted multicolor-blackbody diagnostic, inspired by thin-disk spectral modeling~\cite{Mitsuda:1984,Makishima:1986} not as a ray-traced observed spectrum.


These proxies test if the shifts in $x_{\rm ISCO}$, $\widetilde F(x)$ and $\widetilde T(x)$ are still apparent when the leading gravitational-redshift and Doppler factors are folded in. Light bending, higher order images, returning radiation, finite disk thickness, and coronal transfer are not included. They are essential for precision spectroscopy, but not for the limited comparative diagnostics pursued here.  In addition, for the spectral calculations discussed below, we use
$x_{\rm out}=20$, a radial spacing $\Delta x\simeq0.01$,
and an azimuthal spacing $\Delta\phi=\pi/80$. The quoted
spectral maxima were independently refined using
$\Delta x\simeq0.002$ and a frequency spacing
$\Delta\widetilde{\nu}=2\times10^{-4}$.

\section{Inner-disk thermal diagnostics motivated by neutrino processes}
\label{sec:neutrinos}

Neutrino emission is a sensitive probe of the temperature attained in  the innermost accretion region. In hyperaccreting black-hole systems, especially those described as potential central engines of gamma-ray bursts, the disk can become hot and dense enough that neutrino cooling can contribute significantly to the local energy balance, and neutrino--antineutrino annihilation can deposit energy above the disk and contribute to the power of a relativistic outflow~\cite{ PophamWoosleyFryer1999,DiMatteoPernaNarayan2002, ChenBeloborodov2007,ZalameaBeloborodov2011,Liu2017}. Besides, the sensitivity of this deposition channel to modified-gravity backgrounds has also been investigated using idealized disk models \cite{Lambiase:2022ywp}. 
These processes are strongly temperature-dependent and thus could be sensitive to the entropy-induced shift of the ISCO and the corresponding change of the Novikov-Thorne temperature profile.

A self-consistent treatment of a neutrino-dominated accretion flow lies beyond the scope of this section, where such a calculation would require the radial density profile, vertical hydrostatic structure, electron
fraction, chemical potentials, neutrino opacity, trapping, degeneracy
corrections and relativistic neutrino transport. Here, none of these quantities are resolved. Instead, we introduce dimensionless temperature moments that measure the propagation of an inner-disk temperature deformation into quantities with increasingly higher thermal sensitivity.

The physical disk temperature obtained in Sec.~\ref{sec:disk} is
\begin{equation}
T_{\rm phys}(x) =T_0\,\widetilde T(x).
\label{eq:Tphys_recall_neutrino}
\end{equation}
We define an effective neutrino temperature by
\begin{equation}
T_{\nu,{\rm phys}}(x) = T_0\,\widetilde T_\nu(x),
\qquad
\widetilde T_\nu(x) = \chi_\nu\,\widetilde T(x),
\label{eq:Tnu_tilde}
\end{equation}
unless otherwise specified, $\chi_\nu=1$. A straightforward mismatch between the effective photon temperature and the temperature defining the neutrino-emitting region is permitted by the parameter $\chi_\nu$. The numerical analysis that follows does not assume any such mismatch.

The inner integration boundary is taken to be the ISCO of each
reconstructed geometry,
$x_{\rm in} = x_{\rm ISCO}(\lambda_G,p).$ 
We then introduce the dimensionless temperature moment
\begin{equation}
\mathcal M_\nu^{(n)}(\lambda_G,p)
= \int_{x_{\rm in}}^{x_{\rm out}}
x^2\, \widetilde T_\nu(x)^n\,dx.
\label{eq:nu_temperature_moment}
\end{equation}
As a simplified volume-like radial weight, the factor $x^2$ is employed. Up to an overall normalization, it corresponds to an emitting region whose characteristic scale height satisfies $H(r)\propto r$, for which $dV\propto rH(r)dr\propto r^2dr$. The vertical structure of an accretion flow dominated by neutrinos is not meant to be replaced by this prescription. Its sole function is to apply the same radial weighting to every configuration, enabling consistent comparison of the temperature response.

The exponents
\begin{equation}
n=4,\qquad n=6,\qquad n=9
\label{eq:nu_benchmark_exponents}
\end{equation}
are benchmark measures of thermal sensitivity, not alternative complete
emissivity models. They are motivated by the characteristic
temperature dependence of several processes encountered in hot
accretion flows:
\begin{align}
n=4 &: \quad \text{thermal }T^4\text{-type benchmark},
\label{eq:n4_benchmark}
\\
n=6 &: \quad \text{charged-current or Urca-like temperature dependence},
\label{eq:n6_benchmark}
\\
n=9 &: \quad \text{pair-process or annihilation-like temperature sensitivity}.
\label{eq:n9_benchmark}
\end{align}
The $n=6$ case reflects the leading $T^6$ dependence often encountered
in charged-current cooling rates after suppressing the density and
composition factors. The $n=9$ case probes the much stronger thermal
dependence associated with pair processes and simplified
neutrino--antineutrino annihilation kernels~\cite{Itoh:1996,
ZalameaBeloborodov2011,Liu2017}. These moments should not be interpreted
as mutually normalized physical luminosities, since the corresponding
microscopic rates carry different dimensional coefficients and
additional dependences on density, composition and geometry.

All comparisons are performed at fixed $T_0$, $\dot M$, $r_h$ and
$\chi_\nu$. The dimensional factor $T_0^n$ therefore cancels between
the reconstructed and Schwarzschild configurations. The quantities
defined in Eq.~\eqref{eq:nu_temperature_moment} measure only the
relative change induced by $\lambda_G$ and $p$.

For a thermal Fermi--Dirac distribution with vanishing chemical
potential, the mean neutrino energy is
$\langle E_\nu\rangle=3.151\,k_{\rm B}T_\nu$. This motivates 
the dimensionless temperature-weighted mean energy

%
\begin{equation}
\left\langle\widetilde E_\nu\right\rangle_n
\equiv
\frac{\left\langle E_\nu\right\rangle_n}{k_{\rm B}T_0}
= \frac{ \displaystyle
\int_{x_{\rm in}}^{x_{\rm out}} x^2\, \widetilde T_\nu(x)^n\,
3.151\,\widetilde T_\nu(x)\,dx}{ \displaystyle \int_{x_{\rm in}}^{x_{\rm out}}
x^2\,\widetilde T_\nu(x)^n\,dx }.
\label{eq:mean_Enu}
\end{equation}
The corresponding number-weighted moment is
\begin{equation}
\mathcal N_\nu^{(n)}
=
\frac{\mathcal M_\nu^{(n)}}
{\left\langle\widetilde E_\nu\right\rangle_n}.
\label{eq:nu_number_moment}
\end{equation}
Such a quantity distinguishes between changes in the integrated temperature moment and changes in its characteristic energy scale. Since the weak-interaction coefficients and dimensional normalization have not been added, it is not a physical neutrino number-emission rate.

To probe the stronger concentration of annihilation-related processes
toward the inner disk, we introduce the heuristic kernel
\begin{equation}
\mathcal W_{\nu\bar\nu}(x)
=\left[ 1- \sqrt{ 1-\left(\frac{x_{\rm in}}{x}\right)^2}
\right]^2, \qquad x\geq x_{\rm in}.
\label{eq:annihilation_weight}
\end{equation}
It satisfies
\begin{equation}
\mathcal W_{\nu\bar\nu}(x_{\rm in})=1,
\qquad
\mathcal W_{\nu\bar\nu}(x)
\sim
\frac{x_{\rm in}^4}{4x^4}
\quad
\text{for}
\quad
x\gg x_{\rm in}.
\label{eq:annihilation_weight_limits}
\end{equation}
Therefore, the kernel  emphasizes the region near the inner edge and
suppresses the outer disk. It is motivated by the larger collision
angles and solid angles available close to the central object. 
The corresponding annihilation-weighted temperature moment is
\begin{equation}
\mathcal Q_{\nu\bar\nu}^{(n)}(\lambda_G,p)
= \int_{x_{\rm in}}^{x_{\rm out}}
x^2\, \widetilde T_\nu(x)^n\, \mathcal W_{\nu\bar\nu}(x)\,dx.
\label{eq:Qnunu}
\end{equation}
Equation~\eqref{eq:Qnunu} does not contain any extra redshift power. The simultaneous transformation of neutrino energies, arrival rates, intensities, interaction angles, and weak cross sections would be necessary to measure the physical annihilation rate at infinity.
 Recent relativistic calculations in deformed black-hole backgrounds
provide explicit examples of such a deposition treatment, including
the local temperature-redshift relation and a geodesic angular factor
\cite{Becar:2026doz}.
Henceforth, the redshifted photon spectrum discussed in
Sec.~\ref{subsec:redshift_proxy} addresses a different observational
question and is therefore treated separately. The integrated quantities used below are normalized to their
Schwarzschild counterparts at the same value of $p$:
\begin{equation}
\mathcal R_\nu^{(n)}
= \frac{ \mathcal M_\nu^{(n)}(\lambda_G,p)
}{ \mathcal M_\nu^{(n)}(0,p) },  \qquad \mathcal R_{\nu\bar\nu}^{(n)}
= \frac{ \mathcal Q_{\nu\bar\nu}^{(n)}(\lambda_G,p)
}{\mathcal Q_{\nu\bar\nu}^{(n)}(0,p)}.
\label{eq:neutrino_ratios}
\end{equation}
Values larger than unity indicate an enhancement of the corresponding
temperature-sensitive diagnostic, whereas values below unity indicate
a suppression.

The integrated thermal moments are evaluated up to
$x_{\rm out}=20$ using a composite trapezoidal rule with
$\Delta x=0.05$. Convergence was checked against a reference
calculation with $\Delta x=0.025$ and also monitored for
$\Delta x=0.25$ and $0.10$. Over the representative
$(p,\lambda_G)$ configurations, the maximum relative
differences at the adopted resolution are
$4.7\times10^{-6}$ for the $n=9$ temperature moment and
$8.6\times10^{-6}$ for the annihilation-weighted moment.

For the local comparison, we define
\begin{equation}
\mathcal R_{\nu,{\rm loc}}^{(n)}(x)
= \left[ \frac{ \widetilde T_\nu(x;\lambda_G,p)
}{ \widetilde T_\nu(x;0,p) } \right]^n. \label{eq:local_neutrino_ratio}
\end{equation}
This ratio is evaluated only on a common radial domain where both disk
solutions are physically defined and the Schwarzschild reference flux
is not arbitrarily close to zero.

%

The quantities introduced in this section retain the leading
temperature sensitivity expected from neutrino-emitting processes, but neglect the back-reaction of the neutrino cooling on the disk structure. It should be considered as a temperature diagnostic of the strong-field region, rather than a substitute for NDAF or neutrino-transport calculations.

\section{Strong-field signatures of the thermodynamic reconstruction}
\label{sec:numerical_results}

We now follow the effect of the entropy-induced horizon response from
the orbital structure to the disk emission, the redshifted spectrum,
and the neutrino-sensitive thermal measures. We use the representative values
$\lambda_G=-0.2,0,0.2$  to display the key trends, while
$p=2,3,4$ is varied to test how much such response depends on the
radial localization of the reconstructed correction. In all cases investigated
below, the condition $\lambda_G>-1$ is satisfied and the geometry admits
both a photon sphere and a physical ISCO. 

Two additional quantities are introduced to describe the radial
concentration of the radiative output. The first is the inner-disk
fraction
\begin{equation}
\mathcal C_{20}
=\frac{\widetilde{\mathcal L}_\infty(<20)}
{\widetilde{\mathcal L}_\infty(\infty)} ,
\label{eq:C20_definition_results}
\end{equation}
which measures the fraction of the energy-at-infinity luminosity
generated inside the common scale $x=20$. The second is the
half-luminosity radius $x_{50}$, defined by 
\begin{equation}
\widetilde{\mathcal L}_\infty(<x_{50})
= \frac{1}{2}\,
\widetilde{\mathcal L}_\infty(\infty).
\label{eq:x50_definition_results}
\end{equation}
Since the ideal Novikov--Thorne energy balance gives $\widetilde{\mathcal L}_\infty(\infty)=\eta_{\rm NT}$, these two quantities measure the radial redistribution of the emission separately from the total radiative efficiency.

Table~\ref{tab:compact_observables} summarizes representative compact
and thin-disk quantities for $\lambda_G=0,\pm0.2$ and $p=2,3,4$. 

\begin{table*}[t]
\caption{\footnotesize\itshape
Representative compact and thin-disk quantities for
$\lambda_G=0,\pm0.2$ and $p=2,3,4$. The shadow scale is normalized by
its Schwarzschild value. The quantities $\mathcal C_{20}$ and $x_{50}$
measure the radial concentration of the energy-at-infinity luminosity.}
\label{tab:compact_observables}

\centering
\setlength{\tabcolsep}{4pt}
\renewcommand{\arraystretch}{1.15}

\begin{tabular*}{\textwidth}{
@{\extracolsep{\fill}}
c c c c c c c c
}
\hline\hline
$p$
& $\lambda_G$
& $x_{\rm ph}$
& $\widetilde b_{\rm ph}/\widetilde b_{\rm ph}^{(0)}$
& $x_{\rm ISCO}$
& $\eta_{\rm NT}$
& $\mathcal C_{20}$
& $x_{50}$
\\
\hline
2 & $-0.2$ & 1.5499 & 1.0460 & 3.1910 & 0.05374 & 0.5632 & 16.6076 \\
2 & $0$    & 1.5000 & 1.0000 & 3.0000 & 0.05719 & 0.5823 & 15.6388 \\
2 & $0.2$  & 1.4605 & 0.9572 & 2.8377 & 0.06065 & 0.5992 & 14.8034 \\
\hline
3 & $-0.2$ & 1.5469 & 1.0296 & 3.1747 & 0.05399 & 0.5634 & 16.5909 \\
3 & $0$    & 1.5000 & 1.0000 & 3.0000 & 0.05719 & 0.5823 & 15.6388 \\
3 & $0.2$  & 1.4584 & 0.9704 & 2.8203 & 0.06103 & 0.6028 & 14.6345 \\
\hline
4 & $-0.2$ & 1.5398 & 1.0193 & 3.1130 & 0.05540 & 0.5714 & 16.1916 \\
4 & $0$    & 1.5000 & 1.0000 & 3.0000 & 0.05719 & 0.5823 & 15.6388 \\
4 & $0.2$  & 1.4612 & 0.9798 & 2.8751 & 0.05932 & 0.5946 & 15.0254 \\
\hline\hline
\end{tabular*}
\end{table*}
 Within the power-law localization family considered here, the representative configurations exhibit a common ordering: for $\lambda_G>0$, the photon sphere and
ISCO move inward, the critical shadow scale decreases, and the
Novikov--Thorne efficiency increases, whereas $\lambda_G<0$ reverses
these trends. At $p=2$, increasing
$\lambda_G$ from $0$ to $0.2$ shifts the ISCO from $x_{\rm ISCO}=3$ to
$x_{\rm ISCO}\simeq2.8377$, increases the efficiency by about $6\%$, and
decreases the shadow scale by about $4.3\%$. For $\lambda_G=-0.2$, the
ISCO moves outward by about $6.4\%$ and the efficiency is reduced by a
similar relative amount.

Moreover, within the power-law family, the dependence on $p$ is moderate for the compact geodesic quantities
and is not strictly monotonic. Among the representative cases, the
largest positive efficiency shift occurs at $p=3$, where
$x_{\rm ISCO}\simeq2.8203$ and $\eta_{\rm NT}\simeq0.0610$ for
$\lambda_G=0.2$. Thus, within the power-law family and over the parameter range examined
here, varying $p$ mainly changes the magnitude of the response, while
the same $\lambda_G$-ordering is preserved for $p=2,3,4$.

\begin{figure}[!ht]
\centering
\includegraphics[width=0.42\textwidth]
{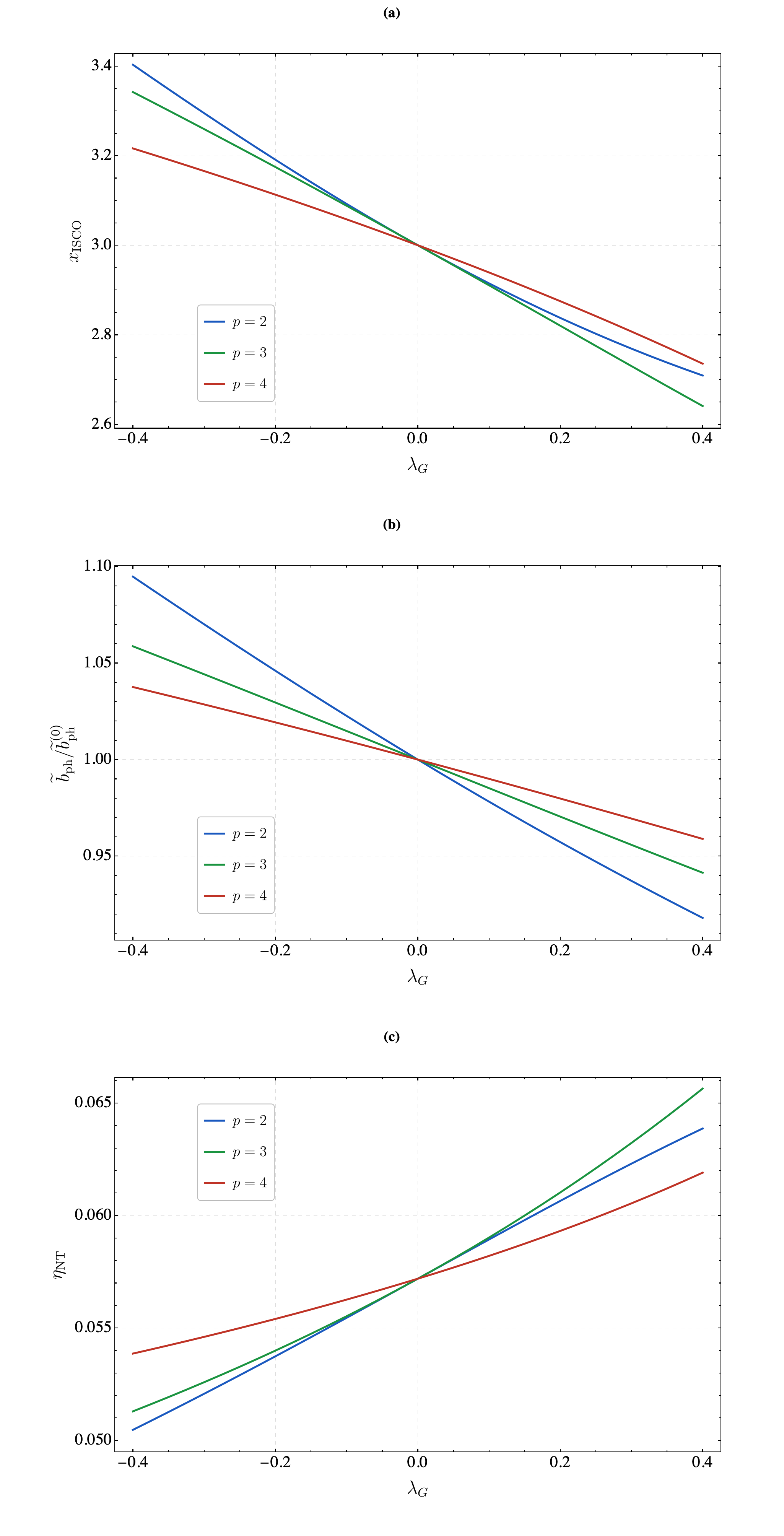}
\caption{\footnotesize{\it Compact strong-field quantities as functions of
$\lambda_G$ for $p=2,3,4$. From top to bottom: the ISCO radius,
the critical shadow scale normalized by its Schwarzschild value, and
the Novikov--Thorne radiative efficiency. 
 Within the power-law reconstruction, positive $\lambda_G$ moves the
ISCO inward, reduces the shadow scale, and raises the efficiency;
negative $\lambda_G$ produces the opposite response.  The dependence
on distinct exterior localization functions is quantified separately
in Table~\ref{tab:profile_robustness}.
}}
\label{fig:compact_observables}
\end{figure}

Figure~\ref{fig:compact_observables} reveals the same behavior in the continuous range of $\lambda_G$. All curves go to the Schwarzschild point at $\lambda_G=0$. The separation between the $p=2,3,4$ profiles is small near the Schwarzschild limit and it becomes more visible as the deformation increases.

\begin{figure}[t]
\centering
\includegraphics[width=\columnwidth]
{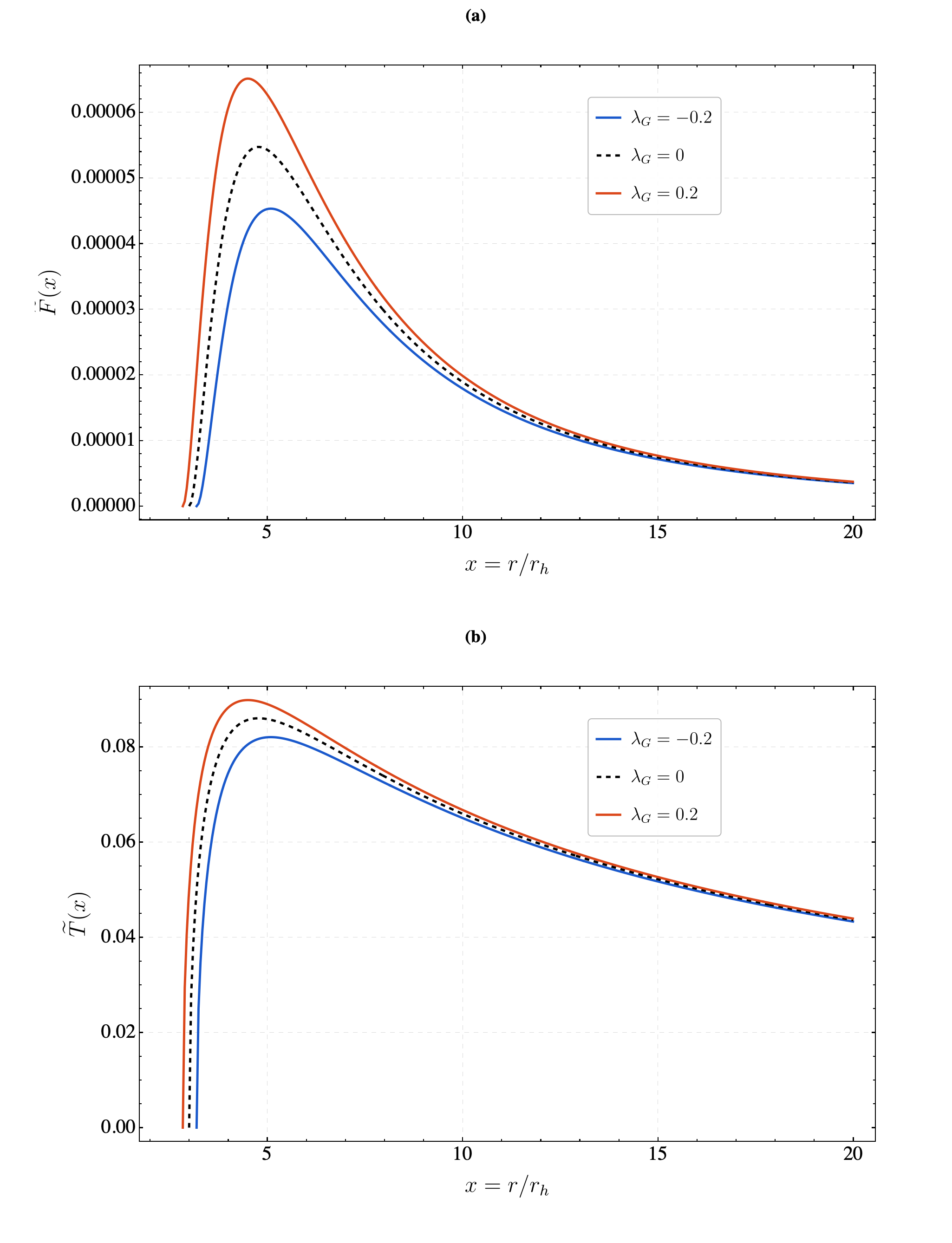}
\caption{Dimensionless Novikov--Thorne flux and effective temperature
for $p=2$ and $\lambda_G=-0.2,0,0.2$. Positive $\lambda_G$ moves the
inner edge inward, increases the peak flux, and raises the maximum
temperature. The profiles approach one another at large radius as the
geometry returns to its Schwarzschild asymptotics.}
\label{fig:disk_profiles}
\end{figure}

The change in the orbit propagates directly into the disk profiles. For $p=2$ the maximum of the dimensionless Novikov--Thorne flux changes from $\approx 4.53\times10^{-5}$ at $\lambda_G=-0.2$ to $5.47\times10^{-5}$ at $\lambda_G=0$ and $6.51\times10^{-5}$ at $\lambda_G=0.2$ (Fig.~\ref{fig:disk_profiles}). This corresponds to a suppression of about $17\%$ for negative $\lambda_G$ and an enhancement of about $19\%$ for positive $\lambda_G$ compared to the Schwarzschild case.  The flux peak also moves inward from $x\simeq5.09$
to $x\simeq4.50$ as $\lambda_G$ changes from $-0.2$ to $0.2$.

In addition, the corresponding change in the maximum temperature is more modest, as expected from the fourth-root relation $\widetilde T=\widetilde F^{1/4}$. The representative values are $\widetilde T_{\max}\simeq0.0821,0.0860,0.0898$ for $\lambda_G=-0.2,0,0.2$, respectively. Thus the deformation is most visible in the flux amplitude and in the location of the inner-disk peak, while the temperature varies more smoothly. At larger radius, the profiles become close to each other, in agreement with the fast decay of the reconstructed correction and the common Newtonian limit.

\begin{figure}[!ht]
\centering
\includegraphics[width=\columnwidth]
{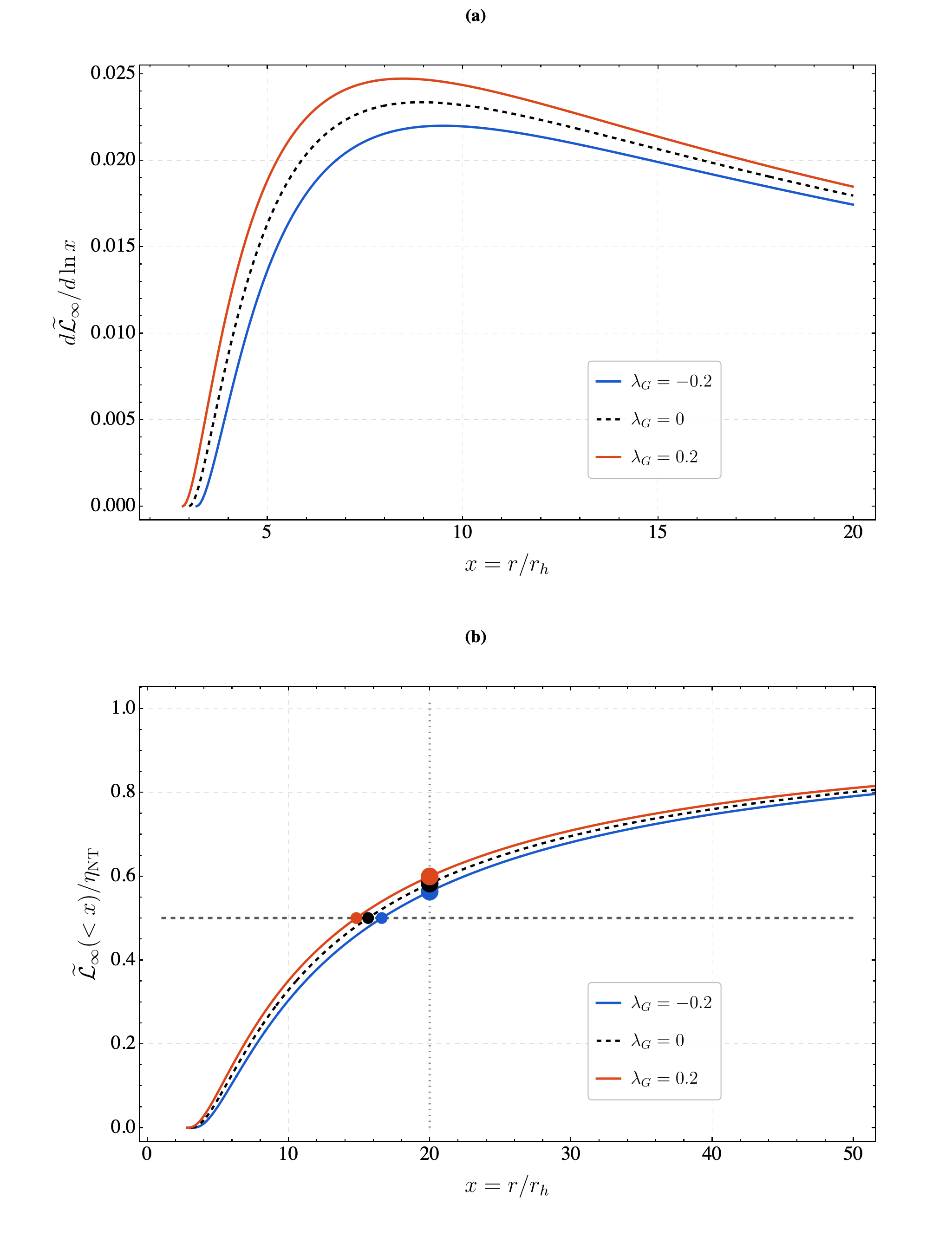}
\caption{Radial distribution of the conserved energy radiated to
infinity for $p=2$. Top:
$d\widetilde{\mathcal L}_\infty/d\ln x$. Bottom:
$\widetilde{\mathcal L}_\infty(<x)/\eta_{\rm NT}$. The horizontal
line identifies the half-luminosity level, the smaller markers locate
$x_{50}$, and the markers at $x=20$ show $\mathcal C_{20}$.
Positive $\lambda_G$ shifts the radiative output toward the inner
disk.}
\label{fig:energy_distribution}
\end{figure}

A clearer picture of the reorganization of the radiative output is provided by the energy-at-infinity distribution. For $p=2$, the upper panel of Fig.~\ref{fig:energy_distribution} displays $d\widetilde{\mathcal L}_\infty/d\ln x$. Its maximum is found close to $x\simeq9.49$ for $\lambda_G=-0.2$, close to $x\simeq8.95$ in the Schwarzschild case, and close to $x\simeq8.49$ for $\lambda_G=0.2$.
Concurrently, the peak value rises. Therefore, a positive horizon response shifts the release of conserved Killing energy toward smaller radii in addition to increasing available efficiency.

This concentration is transparent due to the cumulative curves in the lower panel. For $p=2$, $x_{50}$ falls from $16.61$ to $14.80$, while the fraction $\mathcal C_{20}$ rises from $0.5632$ at $\lambda_G=-0.2$ to $0.5992$ at $\lambda_G=0.2$. Since $\eta_{\rm NT}$ normalizes the cumulative profile, this trend cannot be reduced to the change in the total radiative efficiency. It displays a true radial redistribution of the energy released. The strongest concentration among the representative cases is found for $p=3$ and $\lambda_G=0.2$, with $\mathcal C_{20}\simeq0.6028$ and $x_{50}\simeq14.63$.

\begin{figure}[!ht]
\centering
\includegraphics[width=\columnwidth]
{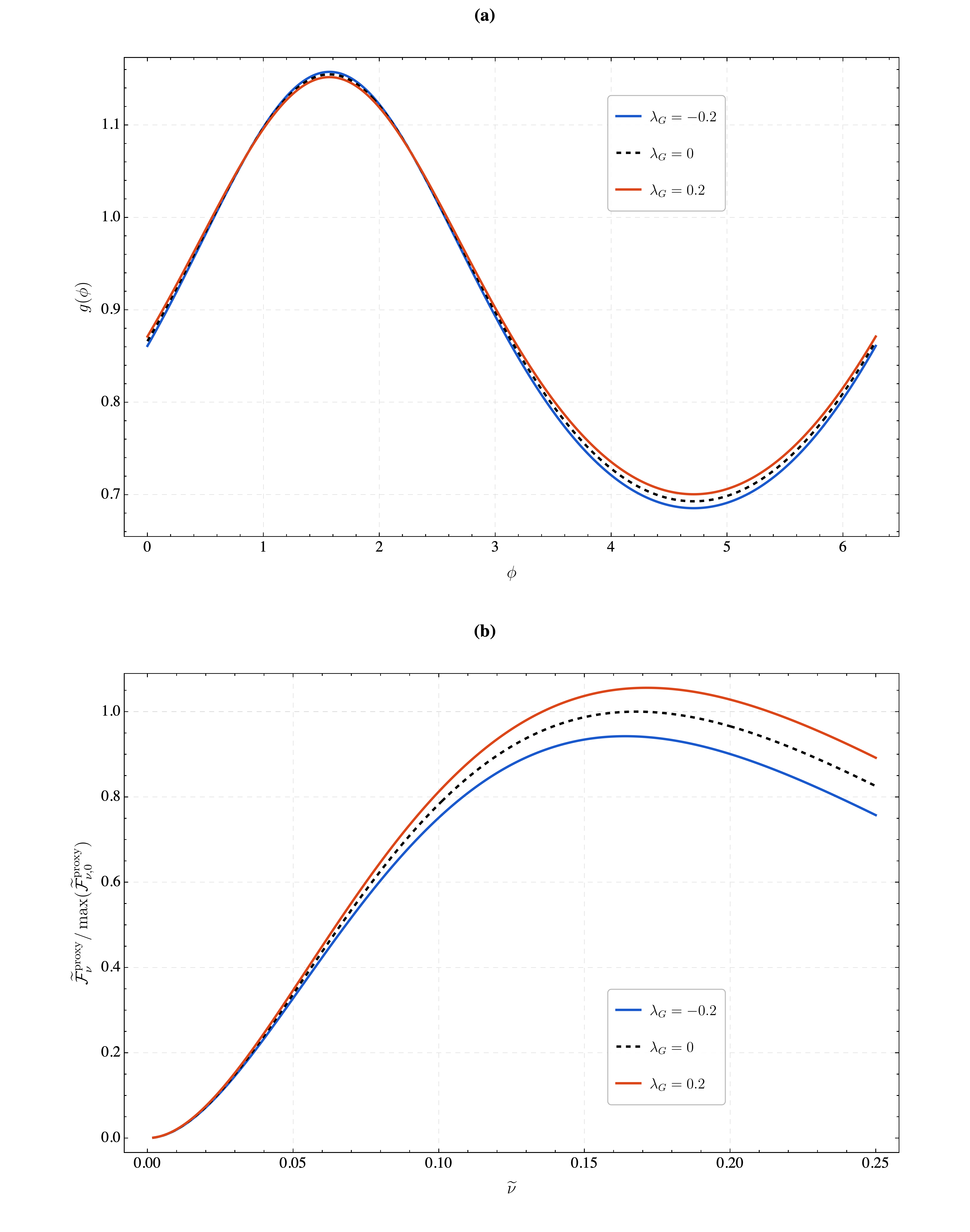}
\caption{\footnotesize{\it  No-bending redshift and spectral diagnostics for $p=2$
and $i=60^\circ$. Top: azimuthal redshift factor at $x=6$.
Bottom: redshifted multicolor-blackbody spectrum integrated up to
$x_{\rm out}=20$ and normalized by the Schwarzschild peak.
The deformation modifies the spectral amplitude more strongly than
the position of the peak.}}
\label{fig:redshift_spectrum}
\end{figure}

When the leading gravitational and Doppler factors are taken into account, Figure~\ref{fig:redshift_spectrum} looks at whether the same ordering remains. For $p=2$ and $i=60^\circ$, the redshift factor $g(\phi)$ is displayed at $x=6$. The fact that the three curves stay near to one another suggests that the local frequency-shift factor's direct change is minimal at this radius. Therefore, rather than the direct variation of $g$, the modified temperature and flux profiles are primarily responsible for the larger change in the spectral amplitude.

 The redshifted multicolor-blackbody spectra are integrated up to
$x_{\rm out}=20$ and normalized by the maximum of the Schwarzschild
spectrum. For $\lambda_G=-0.2,0,0.2$, the normalized peak amplitudes
are approximately $0.9424$, $1$, and $1.0559$, while the corresponding
peak frequencies are $\widetilde{\nu}_{\rm peak}\simeq0.1640$,
$0.1678$, and $0.1714$, respectively. Thus, within the no-bending
construction, the deformation affects the spectral amplitude more
strongly than the position of the peak.

\begin{figure}[!ht]
\centering
\includegraphics[width=\columnwidth]
{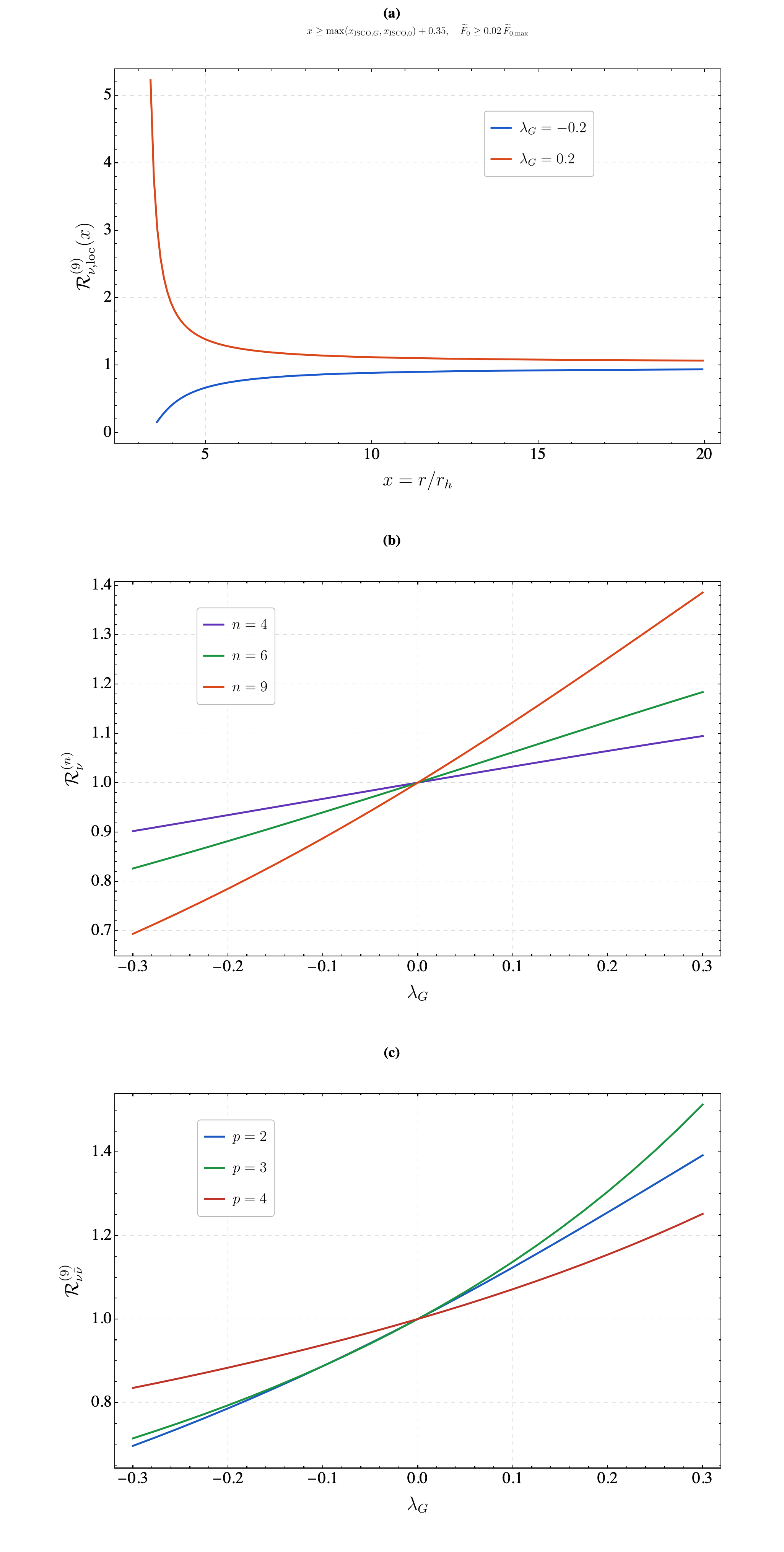}
\caption{Thermal measures motivated by neutrino-emitting processes.
Top: local $n=9$ temperature-moment ratio for $p=2$, evaluated only
where
$x\geq\max(x_{{\rm ISCO},G},x_{{\rm ISCO},0})+0.35$ and
$\widetilde F_0\geq0.02\,\widetilde F_{0,\max}$.
Middle: integrated ratios $\mathcal R_\nu^{(n)}$ for $n=4,6,9$ at
$p=2$. Bottom: annihilation-weighted ratio
$\mathcal R_{\nu\bar\nu}^{(9)}$ for $p=2,3,4$.  Increasing the thermal power amplifies the response to $\lambda_G$,
while the same qualitative ordering is preserved within the plotted
power-law configurations.}
\label{fig:neutrino_thermal_measures}
\end{figure}

Because they involve higher powers of the disk temperature, the thermal measures driven by neutrino processes react more strongly.
The local $n=9$ temperature-moment ratio on the common radial domain defined in Sec.~\ref{sec:neutrinos} is depicted in the upper panel of Fig.~\ref{fig:neutrino_thermal_measures}. Strong inner-disk enhancement is produced by positive $\lambda_G$, whereas the same local measure is suppressed by negative $\lambda_G$. The purpose of this panel is to show how sensitive a $T^9$ weighting is close to the inner edge. The local ratio is not the primary global indicator because it relies on the exact radial cutoff.

The integrated moments are more stable. At $p=2$, the ratios for
$\lambda_G=-0.2$ are approximately
$\mathcal R_\nu^{(4)}=0.934$,
$\mathcal R_\nu^{(6)}=0.881$, and
$\mathcal R_\nu^{(9)}=0.785$. For $\lambda_G=0.2$, they become
$\mathcal R_\nu^{(4)}=1.064$,
$\mathcal R_\nu^{(6)}=1.123$, and
$\mathcal R_\nu^{(9)}=1.252$. Hence the response grows with the thermal
power:
\begin{equation}
\left|\mathcal R_\nu^{(9)}-1\right|
>
\left|\mathcal R_\nu^{(6)}-1\right|
>
\left|\mathcal R_\nu^{(4)}-1\right| .
\label{eq:thermal_sensitivity_hierarchy}
\end{equation}
This hierarchy is the main message of the neutrino-sensitive sector:
 As the temperature exponent rises, the same disk deformation becomes more apparent.

For all three power-law indices $p=2,3,4$, the
annihilation-weighted $n=9$ moment exhibits the same sign pattern.

For $p=2,3,4$, $\mathcal R_{\nu\bar\nu}^{(9)}$ is roughly $1.256$, $1.305$, and $1.154$ at $\lambda_G=0.2$. The corresponding values at $\lambda_G=-0.2$ are roughly $0.786$, $0.793$, and $0.883$.
 As a result, within the power-law family the amplitude of the response
is more sensitive to $p$ than the compact geodesic quantities, while
the same qualitative ordering is preserved for $p=2,3,4$.
 This makes sense: the innermost disk region, where the radial localization of the reconstructed correction is most important, is given greater weight by high-power thermal moments.

 Within the power-law reconstruction used for the detailed disk
analysis, the compact and disk observables display a coherent ordering.
Positive $\lambda_G$ moves the circular-orbit scales inward, reduces
the critical shadow scale, raises the Novikov--Thorne efficiency, and
concentrates the energy release at smaller radii; negative
$\lambda_G$ reverses this sequence.  The comparison of distinct
localization functions in Sec.~\ref{subsec:profile_dependence} shows
that this qualitative ordering is preserved for the exterior profiles
tested here, while its magnitude is not universal.

\FloatBarrier

\section{Conclusion}
\label{sec:conclusion}
We have constructed and analyzed an effective black-hole geometry
thermodynamically constrained by the three-parameter generalized entropy.
 The central idea of the construction is that the entropy deformation should not be understood as a deformation of the radial coordinate or as a preassigned Reissner--Nordstr\"om-like correction. Instead, the generalized entropy enters through the horizon-response factor $\Xi_{\rm h}=\left.\frac{dS_G}{dS}\right|_{S_{\rm h}},$ 
which fixes the generalized temperature through the first law. The effective exterior geometry is then constrained by requiring its surface-gravity temperature to reproduce this thermodynamic temperature. In this way, the horizon area is kept geometrically fixed, while the horizon response is encoded in $\lambda_G=\frac{1}{\Xi_{\rm h}}-1.$
The exterior propagation of this response, however, requires an additional radial localization prescription.

This formalism separates two ingredients that are often conflated: the thermodynamic information fixed at the horizon and the effective choice used to model the exterior radial profile. The Schwarzschild geometry is recovered smoothly at $\lambda_G=0$, the ADM mass remains unchanged for the power-law family with $p\geq2$, and the first law is satisfied by construction. Nevertheless, for $\lambda_G\neq0$ the reconstructed metric is not a vacuum Einstein solution.  
  Its effective source should therefore be interpreted as a
parametrization associated with the thermodynamically constrained
effective geometry rather than as ordinary matter; no underlying
covariant action from which both the generalized entropy and the
metric follow is assumed here.

Within the power-law reconstruction used for the detailed disk analysis, the compact and disk observables display a coherent ordering. Positive $\lambda_G$ moves the circular-orbit scales inward, reduces the critical shadow scale, raises the Novikov--Thorne efficiency, and concentrates the energy release at smaller radii, whereas negative $\lambda_G$ reverses this sequence. For the compact-orbit observables, the comparison of distinct localization functions in Sec.~IV\,A shows that the same qualitative ordering is preserved for the exterior profiles tested here, while the magnitude of the shifts remains profile dependent.

 Within the power-law reconstruction, this orbital ordering propagates consistently into the disk sector. Positive $\lambda_G$ enhances the inner Novikov--Thorne flux and shifts the radiative output toward smaller radii, whereas negative $\lambda_G$ produces the opposite trend. The global energy balance is expressed in terms of the conserved energy radiated to infinity. In the ideal Novikov--Thorne limit,
 $\widetilde{L}_{\infty}(\infty)=\eta_{\rm NT},$ 
which provides a consistency check of the flux calculation. The quantities $C_{20}$ and $x_{50}$ further quantify the associated radial redistribution of the emission.

We also employed a no-bending redshift and inclination proxy to test whether the disk-level ordering remains visible after the leading gravitational and Doppler frequency shifts are included. For the representative power-law configurations considered here, the redshift factor has only a mild dependence on $\lambda_G$. The spectral response is instead driven mainly by the modified Novikov--Thorne flux and effective temperature profile. Consequently, the redshifted multicolor-blackbody proxy changes primarily in normalization, with only a small shift in the peak frequency.

More importantly, the thermal diagnostics motivated by neutrino processes show how strongly the disk response can amplify the effect of the reconstruction. The hierarchy among the $n=4$, $n=6$, and $n=9$ temperature moments demonstrates that the same disk-level modification becomes progressively more pronounced as the temperature exponent increases. The annihilation-weighted moment follows the same qualitative ordering within the power-law family and displays a stronger dependence on $p$, as expected for a quantity dominated by the innermost disk region. These quantities should be regarded as temperature-sensitive diagnostics rather than as substitutes for a self-consistent neutrino-dominated accretion-flow or neutrino-transport calculation.

 The key result is therefore not a universal mapping from generalized entropy to a unique exterior geometry, but a controlled separation of the two ingredients entering the phenomenology. The generalized entropy fixes the horizon-response parameter $\lambda_G$, whereas the radial localization function governs how this response is propagated through the exterior spacetime. Across the three localization profiles tested in Sec.~IV\,A, the qualitative ordering of the compact-orbit observables is preserved, although the magnitude of the shifts remains profile dependent. The detailed accretion, spectral, and thermal calculations are performed for the power-law family only; within that family, the same ordering propagates consistently from the orbital sector to the disk and thermal diagnostics.

Finally, the same entropic formalism may also be applied to apparent horizons in FLRW cosmology. Since such an extension requires its own background analysis and observational treatment, it will be developed separately in future work.

\begin{acknowledgments}
H. El Moumni acknowledges the networking support of COST Action CA22113 (Fundamental challenges in theoretical physics), CA21136 (Addressing observational tensions in cosmology with systematics and fundamental physics), and CA23130 (Bridging high and low energies in search of quantum gravity). He also thanks the Institute of Physics for its support. This work was carried out under the project UIZ 2025 Scientific Research Projects: PRJ-2025-81.
\end{acknowledgments}

\bibliographystyle{apsrev4-2}
\bibliography{references}
\end{document}